\documentclass[twocolumn,epjc3]{svjour3}  

\usepackage{amsmath}
\usepackage{tabularx}
\usepackage{graphicx}
\usepackage{amssymb}
\usepackage{hyperref} 
\usepackage{booktabs}
\usepackage{tikz-feynman,contour}
\usepackage{color}
\usepackage{colortbl}
\usepackage{soul}
\sethlcolor{yellow}

\newcommand{\be}{\begin{equation}}
\newcommand{\ee}{\end{equation}}
\newcommand{\bea}{\begin{eqnarray}}
\newcommand{\eea}{\end{eqnarray}}

\smartqed  

\RequirePackage{graphicx}
\journalname{Eur. Phys. J. C}
\begin{document}

\title{Holographic QCD and 
the muon anomalous magnetic moment\thanksref{t1} 
}


\author{Josef Leutgeb\thanksref{jl,addr1}
        \and
        Jonas Mager\thanksref{jm,addr1}
        \and
        Anton Rebhan\thanksref{ar,addr1}
}

\thankstext{t1}{Preprint of the article published in \url{https://doi.org/10.1140/epjc/s10052-021-09780-8}}

\thankstext{jl}{e-mail: josef.leutgeb@tuwien.ac.at}
\thankstext{jm}{e-mail: jonas.mager@tuwien.ac.at}
\thankstext{ar}{e-mail: anton.rebhan@tuwien.ac.at}


\institute{Institut f\"ur Theoretische Physik, Technische Universit\"at Wien,
        Wiedner Hauptstrasse 8-10, A-1040 Vienna, Austria \label{addr1}
}

\date{Received: date / Accepted: date}

\maketitle

\begin{abstract}
We review the recent progress made in using holographic QCD to study
hadronic contributions to the anomalous magnetic moment of the muon,
in particular the hadronic light-by-light scattering contribution, where
the short-distance constraints associated with the axial anomaly are notoriously
difficult to satisfy in hadronic models. This requires
the summation of an infinite tower of axial vector mesons, which is naturally
present in holographic QCD models, and indeed takes care of the longitudinal
short-distance constraint due to Melnikov and Vainshtein. Numerically the results
of simple hard-wall holographic QCD models point to larger contributions from axial vector mesons than
assumed previously, while the predicted contributions from pseudo-Goldstone bosons agree
nicely with data-driven approaches.

\end{abstract}

\section{Introduction}
\label{intro}

\begin{figure}
\includegraphics[width=0.5\textwidth]{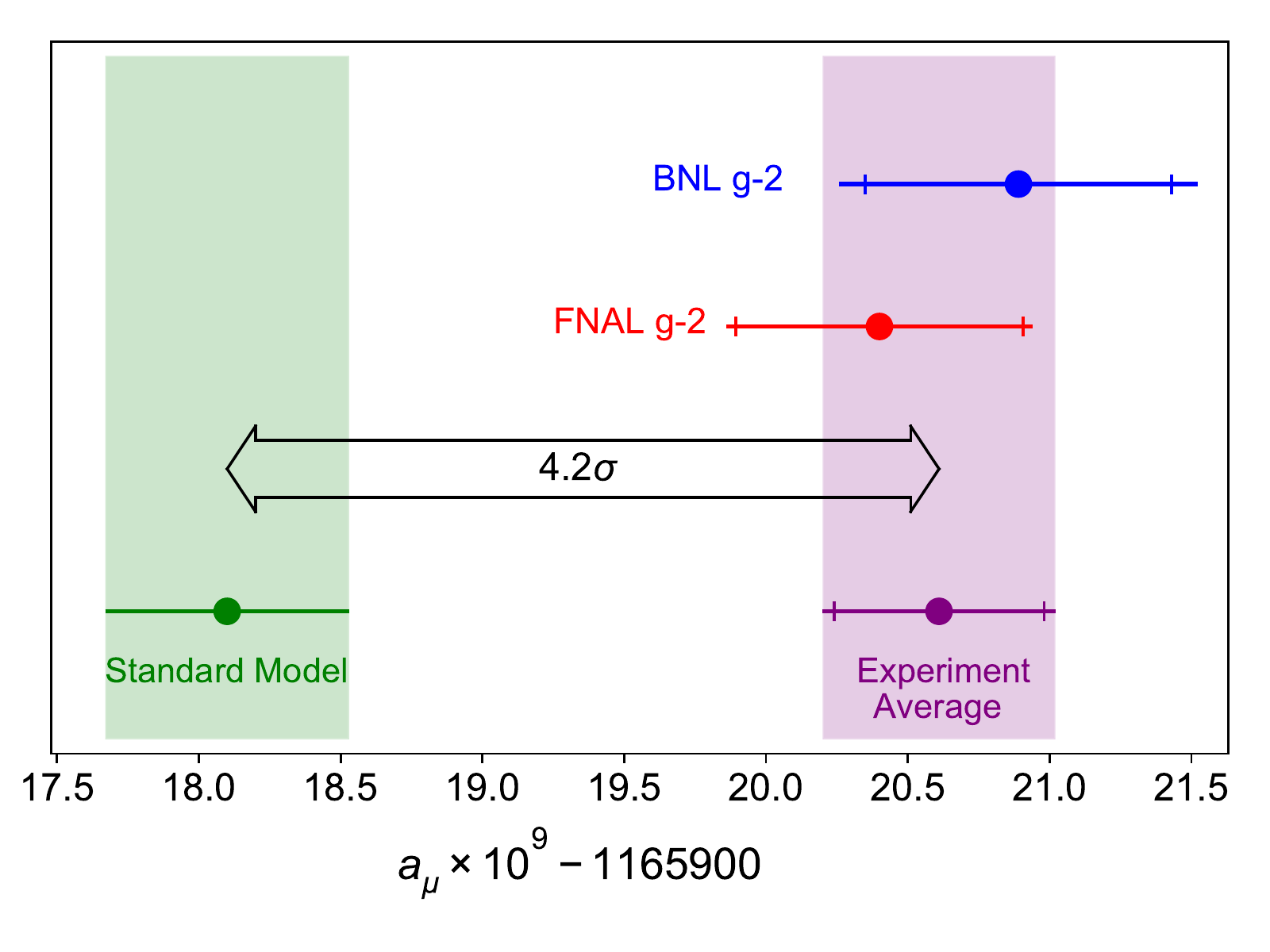}
\caption{Current status of the discrepancy between experiment and SM prediction for the muon anomalous magnetic moment (figure taken from 
\cite{Muong-2:2021ojo})
}
\label{fig:amu2021}
\end{figure}

There is a long-standing discrepancy between the
best theoretical predictions of the anomalous magnetic moment of the muon $a_\mu=(g-2)_\mu/2$ \cite{Jegerlehner:2017gek}
and its experimental value, which was first obtained with
sufficiently high accuracy by the E821/BNL measurement \cite{Muong-2:2006rrc}.
Currently the Standard Model (SM) prediction is at 
\be
a_\mu^{\rm SM(WP)}=116\,591\,810(43) \times 10^{-11}
\ee
according to the 2020
White Paper (WP) of the Muon $g-2$ Theory Initiative \cite{Aoyama:2020ynm}.
The new recent result (2021) by
the Muon $g-2$ Collaboration at Fermilab
\cite{Muong-2:2021ojo} confirmed the BNL result within errors,
slightly reducing the discrepancy from 3.7$\sigma$ to 3.3$\sigma$ when
taken on its own, but raising it to 4.2$\sigma$ when these two independent
measurements are combined in the
new experimental average
\be
a_\mu^{\rm exp.}=116\,592\,061(41)\times 10^{-11},
\ee
as shown in Fig.~\ref{fig:amu2021}.

This gives tantalizing hints if not evidence for physics beyond the Standard Model
which will be further pursued by upcoming improvements of the experimental result
by the Fermilab experiment, and in the medium-term future by a completely new experimental
approach at J-PARC \cite{Abe:2019thb}. On the theoretical side, it will be crucial to continue the efforts
to understand and reduce the uncertainties in the SM prediction. With QED \cite{Aoyama:2012wk,Aoyama:2017uqe,Aoyama:2019ryr} and electroweak effects \cite{Czarnecki:2002nt,Gnendiger:2013pva} being sufficiently under control, the focus in this endeavor is entirely on
hadronic contributions
\cite{Melnikov:2003xd,Prades:2009tw,Kurz:2014wya,Colangelo:2014qya,Pauk:2014rta,Davier:2017zfy,Masjuan:2017tvw,Colangelo:2017fiz,Keshavarzi:2018mgv,Colangelo:2018mtw,Hoferichter:2019gzf,Davier:2019can,Keshavarzi:2019abf,Hoferichter:2018dmo,Hoferichter:2018kwz,Gerardin:2019vio,Bijnens:2019ghy,Colangelo:2019lpu,Colangelo:2019uex,Danilkin:2019mhd,Blum:2019ugy,Chao:2021tvp}, which
require nonperturbative input.
The largest such contribution by far, 
\be
a_\mu^{\rm HVP(WP)}=6\,845(40)\times 10^{-11}
\ee
according to \cite{Aoyama:2020ynm}, 
is from hadronic vacuum polarization (HVP), which is tightly constrained by experimental data. However, the estimated error of just $0.6\%$
has been challenged by a recent lattice calculation \cite{Borsanyi:2020mff} that claims
a comparable accuracy but a 2\% higher value, 
which would reduce the discrepancy between theory and experiment
to a mere $1.5\sigma$ (albeit by
giving rise to tensions in other
sectors of the SM \cite{Crivellin:2020zul,Keshavarzi:2020bfy,Colangelo:2020lcg}).
It is hoped that this question will be resolved in the near future through further lattice
calculations by other groups, while the data-driven approach of \cite{Aoyama:2020ynm} expects
improvements through upcoming new experimental results on low-energy hadronic cross sections. 

Once this is settled, 
the second largest uncertainty, which is due to the
hadronic light-by-light scattering (HLBL) contribution \cite{Danilkin:2019mhd}, 
currently estimated as \cite{Aoyama:2020ynm}
\be\label{amuHLBLWP}
a_\mu^\mathrm{HLBL(WP)}=92(18)\times 10^{-11},
\ee
will also
be crucial for improving the theoretical prediction.
In this case, a data-driven approach is more limited, and hadronic models are widely
used to estimate the numerous contributions from various channels.

The largest contribution to (\ref{amuHLBLWP}) comes from the exchanges
of the neutral pseudo-Goldstone bosons $\pi^0,\eta,\eta'$, whose coupling
to photons is governed by the axial anomaly. As pointed out by Melnikov
and Vainshtein (MV) \cite{Melnikov:2003xd}, the short-distance behavior
of the HLBL amplitude is constrained by the non-renormalization theorems
for the axial anomaly, but conventional hadronic models fail to respect
the so-called longitudinal short-distance constraint (LSDC). 
Using a simple ad-hoc model to correct
for this failure, MV estimated the corresponding effects as a positive
contribution $\Delta a_\mu^\mathrm{MV}=23.5\times 10^{-11}$ which
with current input data would actually become $38\times 10^{-11}$.
The WP result (\ref{amuHLBLWP}) instead uses a much smaller estimate based on
a Regge model for an infinite tower of pseudoscalar bosons constructed
such that the LSDC is satisfied \cite{Colangelo:2019lpu,Colangelo:2019uex},
which was however criticized by MV in \cite{Melnikov:2019xkq}.

In this brief review, we shall describe the recent progress that has been obtained
through holographic QCD and how it helped to clarify
this particular controversy by providing the first hadronic models
where the LSDC constraint can be naturally satisfied in a way that
is consistent with the chiral limit where excited pseudoscalars decouple
from the axial current and thus from the axial anomaly \cite{Leutgeb:2019gbz,Cappiello:2019hwh}. This is brought about
by summing the contributions from the infinite tower of axial-vector mesons
that necessarily appears in holographic QCD, yielding a result
that is larger than the one adopted in the WP estimate, but
clearly below the one obtained in the MV model. Moreover, holographic QCD
points to a significantly larger transverse contribution from axial
vector mesons than assumed in the WP estimate, which has
recently been seconded by re-evaluations of such contributions in resonance chiral theory \cite{Masjuan:2020jsf}.

It should be made clear from the start that holographic QCD is 
only a toy model of real QCD, but it is frequently
remarkably successful, also semi-quanti\-tatively, 
with a minimal set of free parameters.
It certainly cannot help to shed light on the current $\sim 2\%$ discrepancy between data-driven and lattice
approaches to hadronic vacuum polarization. Indeed, holographic QCD results
for the leading light-quark HVP contributions deviate from both data-driven and lattice approaches
at the $\gtrsim 15\%$ level \cite{Hong:2009jv,DA-Stadlbauer,LRSinprep}.
However, as we shall review, 
results for transition form factors (TFF) in the axial sector \cite{Grigoryan:2007wn,Grigoryan:2008up,Grigoryan:2008cc,Hong:2009zw,Cappiello:2010uy,Colangelo:2012ipa,Leutgeb:2019zpq,Leutgeb:2019gbz,Cappiello:2019hwh,Leutgeb:2021mpu},
which are the crucial input
in the HLBL contributions to the muon $g-2$, compare quite well with
data-driven results and so even comparatively simple holographic QCD models
appear to
be useful for estimating the ballparks of various
HLBL contributions, in particular as long as other approaches remain even more uncertain. Note that in the WP \cite{Aoyama:2020ynm} the adopted value for the contribution of axial vector mesons carries a 100\% uncertainty.

\section{HLBL contribution to the muon $g-2$ and short distance constraints}

The magnetic moment $g$ of a particle can be measured by scattering it off of an external electromagnetic field. The probability that the particle will change its spin when interacting with a static magnetic field at small momentum transfer is proportional to $g$. Currently the most interesting particle to look at is the muon. In the Standard Model it is an elementary particle which makes the computations easier as opposed to say a complicated hadronic bound state and compared to the electron it is much heavier and therefore effects of internal loops of heavier particles should be more pronounced. The tau lepton would be even more interesting due to its much higher mass, however it decays much too quickly to measure its magnetic moment accurately. The astonishing precision to which the muon $g-2$ can be measured \cite{Muong-2:2021ojo} and its susceptibility to heavier physics makes it a most interesting testing ground for the Standard Model. If the experimental and the theoretical values disagree significantly, then this clearly signals new physics.

In any QFT coupled to a weak external electromagnetic field one obtains for the scattering amplitude 
\begin{equation}
    \mathcal{M} \propto-ie \int d^4x \langle p^{\prime},\sigma^{\prime}|J^{\mu}(x) |p,\sigma \rangle A_{\mu}(x),
\end{equation}
which means we have to look at diagrams with one incoming and one outgoing muon line and in addition an amputated photon line with momentum $q=p-p'\rightarrow 0$.
Electromagnetic and weak processes are under good theoretical control since one can reliably compute contributions using ordinary perturbation theory in powers of $\alpha$ and $G_F m_\mu^2$ \cite{Aoyama:2020ynm}. This method fails when QCD is included due to it being a strongly coupled theory at low energies. 
The largest hadronic contribution to the magnetic moment comes from hadronic vacuum polarization diagrams (see figure \ref{fig:HVP}), which using dispersion theory can be related to the total cross section for $e^- e^+$ annihilation into hadrons.

\begin{figure}
    \centering
    \begin{tikzpicture}
        \begin{feynman}
            \vertex[blob] (a) at (0,0){\contour{white}{}};
            \vertex (b) at (1,0);
            \vertex (c) at (-1,0);
            \vertex (d) at (0,1);
            \vertex (e) at (0,2-0.3);
            \vertex (f) at (2-0.4,-1+0.4);
            \vertex (g) at (-2+0.4,-1+0.4);
            \diagram*{
            (c)--[boson](a)--[boson](b),
            (g)--[fermion](c)--[fermion](d)--[fermion](b)--[fermion](f),
            (d)--[boson](e),
            };
        \end{feynman}
    \end{tikzpicture}
    \caption{HVP contribution to $g-2$}
    \label{fig:HVP}
\end{figure}
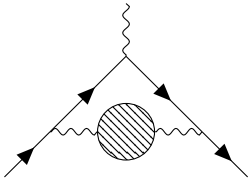

\begin{figure}
    \centering
    \begin{tikzpicture}
        \begin{feynman}
            \vertex (a) at (0,0);
            \vertex (b) at (-1,0);
            \vertex (c) at (1,0);
            \vertex[blob] (d) at (0,1){\contour{white}{$\Pi$}};
            \vertex (e) at (0,2-0.3);
            \vertex (f) at (-2+0.4,0);
            \vertex (g) at (2-0.4,0);
            \diagram*{
            (g)--[anti fermion](c)--[anti fermion](a)--[anti fermion](b)--[anti fermion](f),
            (e)--[boson,edge label'= $q_4$](d),
            (d)--[boson,edge label= $q_3$](c),
            (d)--[boson,edge label'= $q_2$](a),
            (d)--[boson,edge label'= $q_1$](b),
            };
        \end{feynman}
    \end{tikzpicture}
    \caption{HLBL contribution to $g-2$}
    \label{fig:HLBL}
\end{figure}
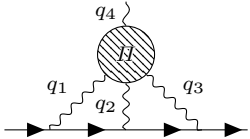

Light-by-light scattering is much less important, however in order to gain the needed precision its contribution has to be analyzed carefully as well. It appears as the set of diagrams as seen in figure \ref{fig:HLBL}. 
The hadronic part of the connected light-by-light scattering subdiagram is to lowest order in $e$ given by a correlator of four electromagnetic currents 
\begin{align}
\label{eq:lbltens}
   & \Pi^{\alpha \beta \mu \nu}(q_1,q_2,q_3):=-i \int d^4x d^4y d^4z e^{-i(q_1x+q_2y+q_3z)} \times \nonumber \\ & \langle \Omega | T \{J^{\alpha}(x) J^{\beta}(y)J^{\mu}(z)J^{\nu}(0)   \} |\Omega \rangle, 
\end{align}
with $J^{\mu}(x)=e \bar{\psi}_0 Q\gamma^{\mu} \psi_0$,
where $\psi_0$ is a multiplet containing the bare quark fields and $Q$ a flavor matrix encoding the charges of the quarks in units of $e$. 

Using Lorentz covariance and gauge invariance one can decompose this tensor into $54$ tensor structures \cite{Colangelo:2015ama} 
\begin{equation}
\label{eq:BTT}
    \Pi^{\alpha \beta \mu \nu}= \sum_i T_i^{\alpha \beta \mu \nu} \Pi_i.
\end{equation}
It is possible to use fewer basis elements, however then for some values of $q_1,q_2,q_3$ one encounters kinematic singularities or zeros.
Inserting this decomposition into the $2$-loop integral of figure \ref{fig:HLBL} to compute $a_{\mu}$ and employing the method of Gegenbauer polynomials first done in \cite{Knecht:2001qf} the final compact formula reads \cite{Colangelo:2015ama}
\bea
\label{eq:integral}
    a_{\mu}&=&\frac{2 \alpha^3}{3 \pi^2} \int_0^{\infty} d Q_1 \int_0^{\infty} d Q_2 \int_{-1}^1 d \tau \sqrt{1-\tau^2} Q_1^3 Q_2^3  \nonumber \\  &&\times\sum_{i=1}^{12}T_i(Q_1,Q_2,\tau) \bar{\Pi}_i(Q_1,Q_2,\tau)\nonumber \\
    &=:& \int_0^\infty dQ_1 \int_0^\infty dQ_2 \int_{-1}^1 d\tau \,\rho_a(Q_1,Q_2,\tau),
\eea
where $T_i$ are known kernel functions and $\bar{\Pi}_i$ are linear combinations of the $\Pi_i$ functions.
The integration region is entirely in the Euclidean regime where no single particle poles or cuts show up.
Before trying to compute this object at low energies using holographic QCD, we first turn to the asymptotic constraints that one can derive from QCD, where the main tool will be the operator product expansion (OPE), used in Euclidean space.

\subsection{Short distance constraints}
Below we will sketch a derivation of the leading-order behavior of $\Pi$ for two different kinematic configurations when the Euclidean momenta $Q_{1,2,3}$ become very large \cite{Melnikov:2003xd}.
In evaluating the HLBL contribution to the muon $g-2$ we are not directly interested in the asymptotic region. The weight functions that appear in the two loop integral of figure \ref{fig:HLBL} fall off quickly beyond $Q \sim 2\, \text{GeV}$. It is however relevant how fast this asymptotic behavior is reached.

In particular one can ask the question of how to satisfy these constraints using 
an effective field theory involving only hadronic degrees of freedom, since
at low energies one cannot use perturbative QCD but requires hadronic models,
unless the entire calculation is done nonperturbatively using lattice QCD
\cite{Blum:2019ugy,Chao:2021tvp}.

An important intermediate result to derive one of the short distance constraints is the OPE of two electromagnetic currents 
\begin{align}
\label{eq:OPE}
    &i\int d^4x e^{-iqx} T\{J_{\mu}(x)J_{\nu}(0)  \}= \nonumber \\
    &2i\varepsilon_{\mu \nu \alpha \beta} \frac{q^{\alpha}}{q^2+i\epsilon} J_5^{\beta}(0)+ \mathcal{O}(\frac{1}{q^2}),
\end{align}
with $J_5^{\beta}$ being $\bar{\psi}_0 Q^2 \gamma_5 \gamma^{\beta}\psi_0$, which holds for large spacelike $q$. The flavor octet part of the RHS is a finite operator and independent of a renormalization scale $M$ which is usually needed when defining composite operators. The flavor singlet part is finite and independent of $M$ up to one loop order but in general it mixes with other operators due to the $U(1)_A$ anomaly. The divergence of the singlet part mixes with the theta term but 
\begin{equation}
    \partial_{\mu} (J_5^{\mu})_M- \frac{\alpha_s(M)}{8 \pi} N_f (\varepsilon^{\mu \nu \rho \sigma}F_{\mu \nu}^a F_{\rho \sigma}^a)_M
\end{equation}
is independent of $M$.

The main strategy to derive the LSDC of \cite{Melnikov:2003xd} is to pick $q_1-q_2$ very large and Euclidean and $q_1+q_2=-q_3$ much smaller (but still larger than the QCD scale). This allows one to insert the OPE \eqref{eq:OPE} into the light-by-light scattering tensor \eqref{eq:lbltens}. In this way the $VVA$ correlator appears in this asymptotic constraint. In real QCD with $m_q \neq 0$ and $N_c=3$ one has to go to large $Q_3^2$ and use asymptotic freedom to obtain
\begin{equation}
\label{eq:MVConstr}
    \lim_{Q_3\rightarrow \infty}\lim_{Q \rightarrow \infty}  Q_3^2 Q^2 \bar{\Pi}_1(Q,Q,Q_3)= -\frac{2}{3 \pi^2},
\end{equation}
which was first derived in \cite{Melnikov:2003xd}. Here we have given the results in terms of the functions appearing in the tensor decomposition \eqref{eq:BTT}.
In the limit of vanishing quark masses, the chiral anomalies allow an exact evaluation of the longitudinal non-flavor-singlet part of the $VVA$, while its transverse part does not contribute to the tensor structure above. In the large-$N_c$ limit 
the $U(1)_A$ anomaly can be ignored for the singlet part.
Equation \eqref{eq:MVConstr} then holds for all $Q_3^2$, and not only asymptotically.

Another short distance constraint follows from considering the symmetric limit $Q_1 \sim Q_2 \sim Q_3$. The one vertex with zero incoming momentum prevents a straightforward use of perturbation theory. Rather one models the vertex by including a non-zero external electromagnetic potential $A_{\mu}$ and then performs an OPE 
\cite{Shifman:1978bx} on the remaining product of $3$ currents. Now condensates can be non-vanishing that would have otherwise been zero by Lorentz invariance like $\langle \bar{\psi}(x) \sigma^{\mu \nu} \psi(x)   \rangle $. The leading behavior is however identical to the perturbative quark loop giving \cite{Melnikov:2003xd,Bijnens:2019ghy}
\begin{equation}
\label{eq:symlim}
    \lim_{Q \rightarrow \infty} Q^4 \bar{\Pi}_1(Q,Q,Q)=- \frac{4}{9 \pi^2}.
\end{equation}

The largeness of the running coupling constant at energies below the QCD scale does not permit the use of perturbation theory in QCD. At low energies it would be desirable to have a different QFT with fields which correspond to single particle states and all the couplings being small at low energy. Demanding for example that $\langle \Omega |\phi (0)|p\rangle=1 $ for a scalar particle of momentum $p$ does not fix the field uniquely. There are many different fields $\phi$ that obey this equation and they all create in some sense this particle, but off-shell they can differ drastically.

Therefore it can happen that two hadronic models, which predict the same current correlators, split up the individual contributions of fields differently. One can for example take one hadronic model and perform a field redefinition to obtain a model that is completely equivalent but has different interactions in the Lagrangian. 
An example would be taking a model with an axial vector field $A_{\mu}$ and a pion field $\pi$ and defining a new axial vector field as $A'_{\mu}=A_{\mu}+{f_{\pi}^3} (\partial_{\mu} \pi)/{\pi^2}$ (the kinetic terms are the same in both formulations).

One-particle intermediate states which are approximately stable show up as poles in the light-by-light scattering amplitude for the right kinematic configurations. These are unambiguously defined since they are on-shell however only for special kinematics.

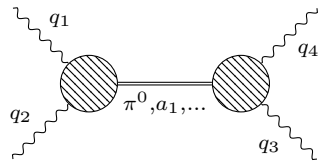
\begin{figure}
    \centering
   \begin{tikzpicture}
     \begin{feynman}
       \vertex[blob] (a) at (0,0) {\contour{white}{}};
        \vertex[blob] (b) at (2,0) {\contour{white}{}};
        \vertex (c) at (-1,1);
        \vertex (d) at (-1,-1);
        \vertex (e) at (3,1);
        \vertex (f) at (3,-1);
        \diagram*{
        (a)--[double,edge label'= {$\pi^0$,$a_1$,...}](b),
        (a)--[boson,edge label'= $q_1$](c),
        (a)--[boson,edge label'= $q_2$](d),
        (b)--[boson,edge label'= $q_4$](e),
        (b)--[boson,edge label'= $q_3$](f)
         };
     \end{feynman}
   \end{tikzpicture}

    \caption{Single-particle intermediate states in the hadronic HLBL tensor}
    \label{fig:pioncr}
\end{figure}

The simplest way that
hadronic degrees of freedom 
appear in the light-by-light scattering tensor 
is through processes where two photon lines connect to a propagator of a charge-neutral hadron which then again splits into two photons  (see figure \ref{fig:pioncr}).
As mentioned before to define an off-shell hadron one needs a corresponding field in the Lagrangian, but as an approximation one can consider an on-shell hadron decaying into two in general virtual photons.

The largest contributions at low energy certainly come from the light charge-neutral (pseudo) Goldstone bosons. For any pseudoscalar particle one defines the transition form factor (TFF) $\mathcal{F}$ via
\begin{align}
    i \int d^4x e^{iq_1x} \langle \Omega | T\{J^{\mu}(x)J^{\nu}(0)  \}| P(p)\rangle \nonumber \\= \varepsilon^{\mu \nu \alpha \beta}(q_1)_{\alpha} (q_2)_{\beta} \mathcal{F}_P(q_1^2,q_2^2)
\end{align}
Axial vector particles can also couple to two virtual photons and one can define form factors analogous to the above equation. Due to the different Lorentz structure there are in general three such form factors. In addition there is the Landau-Yang theorem \cite{Landau:1948kw,Yang:1950rg} which forbids the decay of an axial vector particle into two on-shell photons.
One can also derive asymptotic expressions for these quantities. For simplicity we only give the results for the pions, the other cases can be found in \cite{Hoferichter:2020lap}. The first constraint comes from inserting the OPE \eqref{eq:OPE}, which gives
\begin{equation}
\label{eq:OPEconstr}
   \lim_{Q \rightarrow \infty}Q^2 \mathcal{F}_P(-Q^2,-Q^2) =\frac{2 f_{\pi}}{3}.
\end{equation}
The pion decay constant is defined via 
\begin{equation}
    \langle \Omega| J_5^{a \mu}|P^b(p)\rangle= i \delta^{ab} f_{\pi} p^{\mu}.
\end{equation} 

In the chiral limit for the non-singlet $\pi^0$ the exactness of the anomaly allows to derive
\begin{equation}
\label{eq:lowenergyconstr}
    f_{\pi} \mathcal{F}_{\pi^0}(0,0)=\frac{N_c}{12 \pi}.
\end{equation}
Away from the chiral limit this is not expected to hold anymore; in fact, holographic models predict a correction to the LHS involving 
excited neutral pseudoscalar states as we shall see later.

Finally the single virtual constraint derived in \cite{Brodsky:1981rp,Lepage:1979zb,Lepage:1980fj} reads
\begin{equation}
\label{eq:BrodskyLepconstr}
    \lim_{Q \rightarrow \infty} Q^2 \mathcal{F}_P(-Q^2,0)=2 f_{\pi}.
\end{equation}

The contribution from the neutral pion (with the approximation of an on-shell TFF) to the light-by-light scattering tensor (with one photon being very soft) is 
\begin{equation}
\label{eq:picontr}
    \bar{\Pi}_1 = \frac{\mathcal{F}(q_1^2,q_2^2)\mathcal{F}(q_3^2,0)}{q_3^2-m_{\pi}^2}.
\end{equation}
Using the asymptotic behavior of the TFFs one immediately sees that one cannot reproduce the asymptotic constraints on the HLBL tensor. The same happens for any other single particle intermediate state. 

If one stays within the pole approximation, the only loophole is that one may have to sum over an infinite number of contributions. As we will see the large $N_c$ holographic models naturally provide an infinite number of fields. In particular, the infinite tower of axial vector fields will be responsible for contributing non-zero results for the right-hand sides of the LSDCs (\ref{eq:MVConstr}) and (\ref{eq:symlim}) .

\section{Holographic models}
\label{holmod}

The first concrete realization of the holographic correspondence relating a quantum gauge theory to a higher-dimensional theory with gravity was found by Maldacena in \cite{Maldacena:1997re}, which postulates the full equivalence of $\mathcal{N}=4$ super-Yang-Mills theory in $d=4$ and type IIB closed string theory on a background that is asymptotically $AdS_5\times S^5$. In the limit of infinite 't Hooft coupling, this includes
gauge/gravity duality, i.e., a duality between a strongly coupled quantum field theory and a classical theory of (super-)gravity. 

The duality originates from two different descriptions of branes, the closed string and the open string perspective \cite{Ammon:2015wua}. In the former branes are solitonic objects sourcing the closed string fields and curving the surrounding spacetime.
In the open string perspective branes are described as surfaces on which open strings can end, the low-energy excitations are then described by a $U(N_c)$ supersymmetric gauge theory.

Taking the string coupling constant $g_s \ll 1$ one gets into the regime of classical string theory, and in order to use the classical supergravity approximation, the string length scale $l_s=\sqrt{\alpha'}$ must be small compared to the average length scale of the background curvature. The radius of curvature near the horizon (which is isomorphic to $AdS_5 \times S^5$) for a solution with $N_c$ stacked $D3$ branes  is given by ${R^2}/{\alpha'}=\sqrt{4 \pi g_s N_c}$, and thus $g_s N_c \gg 1$.

Since \cite{Aharony:1999ti} $g_{YM}^2=4 \pi g_s$, the above limit implies that the effective 't Hooft coupling constant $\lambda = N_c g_{YM}^2$ of the large $N_c$ gauge theory is large.

One then takes a low energy limit in which the bulk excitations decouple from the gauge theory. In the supergravity description this same limit zooms in on the near horizon region of the $D3$ branes, whose geometry is $AdS_5\times S^5$ (one also has a background flux that stabilizes the $D3$ branes).

So a weakly coupled semi-classical field theory with gravity in 10 dimensions can be used to make statements about a strongly coupled QFT in 4 dimensions.
This four-dimensional theory is quite different from QCD, since it has conformal symmetry and the maximal amount of supersymmetries but 
breaking these symmetries one can come close to a dual description of large-$N_c$ QCD.

In the ten-dimensional theory gravity is dynamical and the appearance of a background seems irritating at first. What geometry one has deep inside of the space should of course be determined dynamically. Imagine preparing an initial state using a Cauchy hypersurface or letting a bunch of photons start at the timelike asymptotic boundary. Depending on how they start off, they might form a black hole or miss each other entirely. In perturbative string theory however one specifies a fixed background which has to obey certain equations of motion such that the worldsheet theory is a CFT and then studies strings on that background. String theory should ideally be background independent, i.e. one should be able to change to a different background while simultaneously putting all the gravitons into the inital state and still get the same result. Recent computations in a slightly different context \cite{Eberhardt:2021jvj} corroborate this. They show that in their example the perturbative string theory partition function on any fixed background already includes a sum over all semiclassical geometries with the right boundary conditions.

The holographic correspondence also shows this  in a beautiful way. In a QFT one typically considers certain macroscopic states and local excitations thereof. The vacuum state in the QFT corresponds to empty $AdS_5\times S^5$ and a deconfined thermal state to a Euclidean black hole \cite{Witten:1998zw}. For our cases we always look at excitations of the vacuum.

Using the supergravity approximation one can then for example compute the spectrum by looking at normalizable fluctuations around the background and compare them to the Yang-Mills theory. For massless excitations one needs to specify appropriate (reflecting) boundary conditions at the conformal boundary.
One more very useful property of these dualities is how deformations of the four-dimensional theory translate to deformations of the ten-dimensional theory.
The key formula first appeared in \cite{Witten:1998qj} and in terms of the generating functional for connected Greens functions $W_{4D}[J]$ reads
\begin{equation}
\label{eq:holrecipe}
W_{4D}[J]=   S_{\text{grav.}}[\phi(x,z)\rightarrow J(x)],
\end{equation}
where $z$ is the holographic coordinate which goes to zero at the conformal boundary, $J(x)$ is a source coupled to an operator $O(x)$ and $\phi(x,z)$ is said to be the field dual to $O(x)$.
$\phi$ has to satisfy the equations of motion subject to a boundary condition near $z=0$.
The precise statement is a bit more involved, the way $\phi$ behaves near the conformal boundary depends on its mass and whether it is a scalar field, a vector field, etc. 
Thus for every operator in the gauge theory there should be a corresponding field in the supergravity description.

The above formula allows one to compute general correlation functions in the gauge theory by solving classical equations of motion in a gravitational background subject to boundary conditions. 

By now there are a number of holographic models available whose dual theories aim to approximate QCD. Most are ad-hoc bottom-up constructions, where one specifies a set of fields and a background geometry by hand, but there is also one top-down string-theory construction which stands out and provides inspiration for the construction of bottom-up models.
This top-down model was constructed by Sakai and Sugimoto in \cite{Sakai:2004cn,Sakai:2005yt} by embedding probe $D8$ and $\overline{D8}$ branes into the Witten background found in \cite{Witten:1998zw}. We will briefly summarize its construction and main properties before moving on to various bottom-up models.

\subsection{The Witten-Sakai-Sugimoto model}
\label{WSS}

One way to obtain a pure four-dimensional gauge theory from a supersymmetric one is to compactify a five-dimensional theory on a circle $S^1$ with a supersymmetry breaking spin structure (the same one used in thermal partition functions). Type-IIA string theory has stable branes with an odd spacetime dimension which can be put to use for setting up a holographic gauge/string duality.

In the Witten model of pure Yang-Mills theory \cite{Witten:1998zw}, one considers a state with $N_c$ coinciding $D4$-branes, where one spatial direction $x_4$ along the branes is compactified to a circle with radius $M_\mathrm{KK}^{-1}$. On this circle supersymmetry breaking boundary conditions are chosen, generating masses for the fermion (gaugino) fields by the analog of odd Matsubara frequencies and also for the scalar fields because they are no longer protected by gauge symmetry, leaving only $SU(N_c)$ gauge bosons at low energy in the open string perspective. Sakai and Sugimoto in \cite{Sakai:2004cn,Sakai:2005yt} included left and right handed fermions in the fundamental representation by embedding $N_f$ $D8$-$\overline{D8}$-brane pairs which extend in all directions except $x_4$. The $D8$-branes are separated asymptotically from the $\overline{D8}$-branes by a distance $L$ (usually chosen to be maximal), which guarantees that perturbatively strings stretching between the $D8$ and the $\overline{D8}$ branes get a mass. At low energies this system thus contains chiral quarks localized at the intersections of the $D8$ branes with the $D4$ branes and $SU(N_c)$ gauge bosons living in an effectively four-dimensional spacetime. 
One can then again take a limit such that the IIA closed strings in the bulk decouple from the brane excitations.

In the closed string perspective branes source the gravitational fields and thus the $N_c$ $D4$-branes generate a gravitational background with a horizon, while the $D8$-branes are treated in the probe approximation. One embeds the probe branes by minimizing the DBI action and finds that the $D8$ and $\overline{D8}$-branes have to join at the horizon for the simplest case of embeddings, which breaks the initial flavor symmetry $U(N_f)_L\times U(N_f)_R$ spontaneously to its diagonal $U(N_f)_V$, providing a simple geometrical realization of nonabelian chiral symmetry breaking.
The low energy excitations of this system are now the usual type IIA low energy closed string states in the asymptotically flat region far away from the horizon and excitations with arbitrarily high local energy close to the horizon. The latter include strings on the $D8$ branes and excitations of the IIA bulk fields near the horizon. High local energy excitations near the horizon have very small energy for an observer near asymptotic infinity due to the redshift caused by the gravitational field.
Comparing the two different descriptions and taking a decoupling limit once again one can postulate thereby a holographic duality between large $N_c$ QCD with large 't Hooft coupling $\lambda$ and at low energy and the above mentioned near horizon excitations.

Under some assumptions one can perform a Kaluza-Klein compactification of the $D8$-brane action yielding a five-dimensional gravitational action with a certain background geometry which involves (among other fields) two sets of flavor gauge fields whose normalizable modes are interpreted as vector and axial vector meson states in the dual theory, involving also  pseudoscalar bosons as Goldstone modes.  
The geometry is not $AdS_5$ however it also has a timelike conformal boundary like $AdS$. The holographic coordinate ranges from $z=0$ at the conformal boundary to 
a finite value $z_0$ in the infrared, where the circle in $x_4$ smoothly shrinks to zero. 
The flavor gauge fields necessarily obey boundary conditions at $z=z_0$ which break the flavor symmetry $U(N_f)_L\times U(N_f)_R \to U(N_f)_V$. Recall that when one has a gauge symmetry only the global part of it is represented as operators on the Hilbert space and the much larger set of local transformations are only a redundancy of the description. The boundary conditions will of course also break this global part. 
Besides the geometrical realization of chiral symmetry breaking, confinement is brought about by having the holographic coordinate reduced to a finite interval. In accordance with confinement, 
the number of states below a given energy is independent of $N_c$. 

The $D8$-brane action also contains a Chern-Simons term that correctly reproduces the behavior of the QCD partition function when performing a gauge transformation on external flavor gauge fields coupled to the $U(N_f)_L\times U(N_f)_R$ flavor symmetry currents. The flavor anomalies are thus naturally and correctly reproduced.
The $U(1)_A$ anomaly which is a $\frac{1}{N_c}$ effect is also correctly implemented \cite{Sakai:2004cn,Bartolini:2016dbk,Leutgeb:2019lqu} and one can even compute the mass of the $\eta'$ meson using the Witten-Veneziano \cite{Veneziano:1979ec,Witten:1979vv} formula. Baryons naturally show up as instantonic five-dimensional solitons that are dual to Skyrmions in this model \cite{Hata:2007mb}.

A critical short-coming of the Witten-Sakai-Sugi\-moto model is that one cannot take the limit $M_\mathrm{KK}\to\infty$ without leaving the supergravity approximation. At energies much larger than $M_\mathrm{KK}$ the Witten model eventually shows its inherently five-dimensional nature; it can serve as an approximate dual to large-$N_c$ QCD only in the low-energy limit. However, by restricting oneself to zero modes with respect to the extra dimension $x_4$ one can use this model also up to the scale of $M_\mathrm{KK}$ and to some extent even above it.

\subsection{Bottom-up models}
Even before the construction of the (Witten-)Sakai-Sugimoto model (SS), the physics
of chiral symmetry breaking and hadrons was modeled by hand-made bottom-up holographic models, but the lessons learned from string theory help to
understand them and also provide hints for further developments.

We will in the following describe the so-called hard wall (HW) models \cite{Erlich:2005qh,DaRold:2005mxj,Hirn:2005nr}. 
In these models one uses the conformal $AdS_5$ background of the $\mathcal{N}=4$ super-Yang-Mills duality but breaks conformal symmetry by a sharp cut-off in the bulk in order to account for confinement.
In Poincar\'e coordinates where $z=0$ represents the conformal boundary the hard wall is placed at $z=z_0$. One also has two sets of flavor gauge fields $L_M,R_M$ which are holographically dual to the $U(N_f)_L\times U(N_f)_R $ currents in the four-dimensional theory. In some variants also a bifundamental complex scalar field $X_{ij}$ is introduced, whose dual operator is the quark condensate $q^{\dagger}_{Ri} q_{Lj}$ and which permits to introduce finite quark masses through non-normalizable modes.
The action has the usual kinetic terms of the gauge fields and the scalar is minimally coupled to them. One also adds a five-dimensional Chern-Simons term to the action in order to reproduce the flavor anomalies of the dual QFT, thereby implementing the idea of anomaly inflow \cite{Callan:1984sa}.

The first bottom-up models by Erlich et al. \cite{Erlich:2005qh}
and Da Rold and Pomarol \cite{DaRold:2005mxj,DaRold:2005vr} (called HW1 in \cite{Cappiello:2010uy,Leutgeb:2019zpq,Leutgeb:2019gbz,Leutgeb:2021mpu} and in the following) implemented chiral symmetry breaking by choosing a suitable
background solution for the bifundamental scalar $X$, while choosing symmetric infrared boundary conditions for the flavor gauge fields.
Chiral symmetry breaking can however be implemented alternatively, as in the Sakai-Sugimoto model, by specifying asymmetric boundary conditions for the flavor gauge fields at $z_0$.
This is done in the Hirn-Sanz (HW2) model \cite{Hirn:2005nr} model, which refrains from
introducing a bifundamental scalar. The latter is also absent in the Witten-Sakai-Sugimoto model, because there left and right handed fermions are dimensionally separated (quark masses need non-local sources at the boundary and stringy realizations in the bulk \cite{Aharony:2008an,Hashimoto:2008sr}).

In \cite{Domenech:2010aq,Leutgeb:2021mpu} both of these different mechanisms have been applied simultaneously, giving rise to what will be referred to as HW3 model. Normalizable modes of the flavor gauge fields correspond to vector and axial vector particles while normalizable fluctuations of the $X$ field give the scalars and pseudoscalars. In models without the bifundamental scalar $X$ (the HW2 model and the WSS model), the towers of scalars and pseudoscalars is absent, but one can find a massless multiplet of pseudoscalars contained in $ U(x)=\xi_R(x)\xi_L(x)$ with $\xi_{L(R)}=P \exp \big(-i \int_0^{z_0} dz \: L_z(R_z)  \big)$ being Wilson lines. It is important that the IR boundary conditions break the symmetry group down to $U(N_f)_V$ since these make it impossible to gauge $L_z=R_z=0$ everywhere.

The HW models have the attractive feature of involving a minimal set of
parameters which, as we shall see below, permits to fit the most important parameters of low-energy QCD such as $f_\pi$ and $m_\rho$ as well as
certain leading-order pQCD constraints. There have been a number of successful attempts to improve the HW models with their simple AdS background geometry.
Linear confinement and a correct behavior of Regge trajectories in the
high-mass region can be achieved by introducing a nontrivial dilaton which produces a soft rather than a hard cut-off in the so-called soft-wall (SW) models \cite{Ghoroku:2005vt,Karch:2006pv,Kwee:2007dd,Gursoy:2007cb,Gursoy:2007er,Colangelo:2008us,Gherghetta:2009ac,Branz:2010ub,Colangelo:2011xk}, which has similarities with light-front holographic QCD \cite{Brodsky:2014yha},
and also in models where chiral symmetry breaking is described by open-string tachyon condensation
\cite{Casero:2007ae,Iatrakis:2010zf,Iatrakis:2010jb}. Moreover, limitations
of the 't Hooft limit $N_f\ll N_c\to\infty$ can be overcome by
considering back-reactions of flavor fields and thereby covering instead
the Veneziano limit $N_f\sim N_c\to\infty$
\cite{Jarvinen:2011qe}.

\section{Spectra and decay constants in simple HW models}

In this section we consider some of the most basic observables one can compute in the holographic QCD models. For definiteness we will focus on the hard wall models with the additional scalar field $X$. The five-dimensional theory is weakly coupled so masses are given to a good approximation by a classical computation. On a curved spacetime a Fock space is constructed by computing the solutions to the linearized equations of motion and demanding that they are normalizable with respect to a certain inner product (which depends on the type of field one looks at). 
The linearized equations of motion are derived from the Yang-Mills and matter action
\begin{align}
    S&= -\frac{1}{4g_5^2} \int d^4x dz \sqrt{-g} \; \text{tr}\big(|F_L|^2+|F_R|^2\big) \nonumber \\ &+\int d^4x dz\sqrt{-g}\; \text{tr}\big(|DX|^2-M_X^2|X|^2\big).
\end{align}
The metric is taken to be that of Poincar\'e patch $AdS_5$, 
\be
ds^2=z^{-2}(\eta_{\mu\nu}dx^\mu dx^\nu - dz^2),
\ee
with conformal boundary at $z=0$ and a hard-wall cut-off at $z=z_0$
and a mostly-minus signature.
The value of $M_X^2=-3$ is determined by the scaling dimension of the dual
operator of $X$, the bifundamental quark bilinear $ q^{\dagger}_R q_L$. Ref.~\cite{Domenech:2010aq}
has proposed to generalize this to the holographically allowed range
$-4\le M_X^2 \le 0$ and make $M_X$ a parameter that permits more realistic fits
of the hadronic spectrum. In the following we will indicate this modification
by attaching a prime to the name of the various HW models.

Solving the equations of motion for the scalar field $X$, one obtains $X(z)=\frac{1}{2}(M_q z+\Sigma z^3)$ for $M_X^2=-3$, where $M_q$ and $\Sigma$ are proportional to the quark mass matrix and to the chiral condensate, respectively. A  nonzero value of $M_q$ will introduce explicit breaking of chiral symmetry, while $\Sigma$ describes its spontaneous breaking. 
Fluctuations of $X$ involve scalar and pseudoscalar fields; we parametrize the latter,
denoted by $\pi$, through
$X(x,z)= e^{i \pi(x,z)}X(z)e^{-i \pi(x,z)}$.

\begin{table*}[t]
\caption{Masses and decay constants in the axial sector in MeV, compared with experimental masses according to PDG \cite{PDG20} and phenomenological values
for $f_{\pi^*}$ and $F_{a_1}/m_{a_1}$ from \cite{Maltman:2001gc} and \cite{Yang:2007zt}, respectively. (Our definition of $F_{a_1}$ corresponds to $F_A^{a=3}m_A$ and $F_A^{a=3}m_A/\sqrt2$ in \cite{Hoferichter:2020lap,Zanke:2021wiq} and \cite{Yang:2007zt}, respectively.) Fitted values are marked by a star.}
\label{tab:mdc}
\begin{tabular*}{\textwidth}{@{\extracolsep{\fill}}lccccc@{}}
\hline
 & $m_{\pi^0}$ & $m_{\pi^*}$ & $f_{\pi^*}$ & $m_{a_1}$ & $F_{a_1}/m_{a_1}$ \\
\hline
experiment/pheno & 135 & 1300(100) & 2.20(46) & 1230(40) & 168(7) \\
HW2 & 0 & - & - & 1235 & 180 \\
HW2(UV-fit) & 0 & - & - & 1573 & 229 \\
HW1 (chiral) & 0 & 1899 & 0 & 1375 & 177 \\
HW1m & 135$^*$ & 1892 & 1.56 & 1367 & 175 \\
HW1m' & 135$^*$ & 1591 & 1.59 & 1230$^*$ & 148 \\
HW3m & 135$^*$ & 1715 & 1.56 & 1431 & 195 \\
HW3m' & 135$^*$ & 1300$^*$ & 1.92 & 1380 & 186 \\
\hline
\end{tabular*}
\end{table*}

For vector mesons $(L_{\mu}+R_{\mu})/2=V_{\mu}$, the boundary conditions in the infrared are chosen as $F^V_{\mu z}|_{z_0}=0$, which does not interfere with gauging $V_z=0$ for all $z$. The $V_z$ equation of motion then implies that the longitudinal part of $V_{\mu}$ vanishes identically. The eigenvalue equation for the transverse part $\sum_n \psi_n(z)V^{(n)}_{\mu}(x)$ is 
\begin{equation}
    \partial_z[ \frac{1}{z} \partial_z \psi_n(z)  ]+\frac{1}{z} M_n^2 \psi_n(z)=0,
\end{equation}
subject to boundary conditions $\psi_n(0)= \psi_n'(0)=0$. This is solved by Bessel functions 
$\psi_n(z) \propto zJ_1(M_nz)$ with $M_n=\gamma_{0,n}/z_0$ being multiples of the zeros of the Bessel function $J_0$. The location of the hard wall $z_0$ determines the overall mass scale and usually one fits it such that the mass of the rho meson at
approximately $775$ MeV is reproduced by $M_1$.

To compute correlation functions on the QCD side one prescribes boundary values for the fields in the gravity side. For the correlator of vector currents $J^{a \mu}_V$ one turns on boundary values for $V$ and solves the equations of motion. This will lead to a non-normalizable solution which depends on the boundary value. We write the four-dimensional Fourier transform as
\begin{equation}
    \tilde{V}_{\mu}(Q,z)=\mathcal{J}(Q,z)(\tilde{V}_b)_{\mu}(Q)
\end{equation}
where 
$\tilde{V}_b$ is the boundary value and $\mathcal{J}$ is the vector bulk-to-boundary propagator, which in all HW models is given by a simple expression involving Bessel functions only,
\be\label{HWVF}
\mathcal{J}(Q,z)=
Qz \left[ K_1(Qz)+\frac{K_0(Q z_0)}{I_0(Q z_0)}I_1(Q z) \right].
\ee
Using the recipe \eqref{eq:holrecipe} one can then compute the VV correlation function $\Pi_V$ and compare to the OPE result,
\bea\label{PiVas}
\Pi_V(Q^2)&=&-\frac1{g_5^2 Q^2} \left( \frac1z \partial_z \mathcal{J}(Q,z) \right)\Big|_{z\to0}\nonumber\\
&=&-\frac{N_c}{24\pi^2}\ln Q^2,
\eea
which matches the asymptotics perfectly provided $g_5^2=12 \pi^2/N_c$.
The bulk-to-boundary propagator also encodes the respective decay constants since the vector mesons must show up as intermediate states.
For $q^2$ near the mass of a vector meson polology gives
\begin{align}
\int d^4x e^{iqx}    \langle \Omega|T\{J^a_{\mu}(x)J^b_{\nu}(0)\} |\Omega \rangle= \frac{i}{q^2-m^2+i\varepsilon} \nonumber \\ \times \sum_{\lambda} \langle \Omega |J^a_{\mu}(0)|q,\lambda \rangle \langle q,\lambda|J^b_{\nu}(0)|\Omega \rangle \nonumber \\
=\frac{i(\eta_{\mu \nu}-q_{\mu}q_{\nu}/m^2)}{q^2-m^2+i\varepsilon} (F^V)^2 \delta^{ab}
\end{align}
where we used $\langle \Omega | J^a_{\mu}(0)|q,m,b\rangle = F^V \varepsilon_{\mu}(q,\lambda) \delta^{ab}$ and a polarization sum.
In holographic QCD the vector current correlator decomposes into a sum over pole contributions (as should be the case in any large $N_c$ model) and in terms of the radial functions $\psi_n(z)$ the decay constants read $F^V_n= |\psi'(\epsilon)/(g_5 \epsilon)|$ with $z=\epsilon\to0$.

For axial vector mesons the relevant field is $A_M= (L_M-R_M)/2$ and due to the expectation value of $X$ one has an additional contribution to the kinetic terms coming from $|DX|^2$. Depending on which boundary conditions one chooses for the axial vector field in the infrared  one has two possible outcomes. If it is possible to completely gauge $A_z=0$ the $A_z$ equation of motion allows one to express the longitudinal part of $A_{\mu}$ in terms of the $\pi(x,z)$ field contained in $X$. This field then contains an infinite tower of pseudoscalar mesons with a massless multiplet of Goldstone bosons in the chiral limit. This is the case for the so called HW1 model. If the boundary conditions on $A_{\mu}$ in the IR are not like in the vector case, it is in general not possible to have $A_z=0$ everywhere and pion degrees of freedom hide in the Wilson line $P \exp (i \int_0^{z_0} dz A_z)$. The $A_z$ equation of motion will again be responsible for relating the pseudoscalar degrees of freedom. A more convenient gauge in this case is the unitary gauge as discussed in \cite{Domenech:2010aq,Leutgeb:2021mpu}. 

In the first of the above cases spontaneous symmetry breaking only happens through $X$ obtaining a VEV, while in the second case the boundary conditions additionally break the symmetry spontaneously.

While the details of the dynamics are different, in both cases one obtains a tower of pseudoscalar mesons and a tower of transverse axial vector mesons by computing normalizable solutions to the equations of motion. (In the simpler HW2 model, where no
bifundamental scalar $X$ is introduced, since chiral symmetry breaking can be implemented
by boundary conditions alone, only the Goldstone boson appears.)

In Table \ref{tab:mdc}, the results for the first excited pseudoscalar and
the ground-state axial vector mesons are shown and compared with experimental
and phenomenological values, where HW1m and HW3m correspond to models
with isospin symmetric light quark masses chosen such as to reproduce the
mass of the neutral pion $\pi^0$; HW1m' and HW3m' use a tunable value of $M_X^2$
such as to fit the mass of the lightest axial vector meson $a_1$ and the
lightest excited pseudoscalar $\pi(1300)$, respectively. In all models,
$f_\pi$ is set to 92.4 MeV, which in the HW2 model requires to fit
$g_5$ such that the asymptotic constraint (\ref{PiVas}) can no longer be
satisfied fully, but is attained only at the level of 61.6\%.
The model called HW2(UV-fit) in Table \ref{tab:mdc} (termed ``Set 2'' in \cite{Cappiello:2019hwh}) instead keeps $g_5$ as fixed by (\ref{PiVas}), which
with unchanged $f_\pi$ requires a different value of $z_0$, corresponding to
an excessively high value of $m_\rho$ of 987 MeV.

In the HW1/3 models, where $m_\rho$ can be fixed at 775.5 MeV, the
rho meson decay constant has the value $F_\rho^{1/2}=329$ MeV, which
is only 5\% below the phenomenological value 346.2(1.4) MeV of Ref.~\cite{Donoghue:1992dd}. The HW2 value is about 7\% above at 372 MeV, while the HW2(UV-fit) result is at 419 MeV. 

It turns out that the mass of the lightest axial vector meson is
surprisingly well reproduced by the simple HW2 model, while it comes out
at somewhat too large, by 10-20\% in the HW1/HW3 models, and still higher in the HW3 models.
In the HW1m' model, it is in fact possible to match the mass of $a_1$, which
also reduces the mass of the first excited pseudoscalar, which is otherwise
too high in the HW1 and HW3 models by $30-50\%$. While the HW1m' cannot be tuned to
fit the mass of $\pi(1300)$, this is possible in the HW3m' model (which was
the motivation for this particular generalization of the HW model in
\cite{Domenech:2010aq}), but the mass of $a_1$ comes out somewhat too large.

In the SW model, which has more realistic Regge trajectories for the infinite tower of excited states \cite{Ghoroku:2005vt,Karch:2006pv,Kwee:2007dd,Gursoy:2007er,Gursoy:2007er}, it turns out that the
mass of the lightest axial vector comes out as 1674 MeV \cite{Leutgeb:2019gbz}
and thus deviates more strongly from experiment.


Also given in Table \ref{tab:mdc} are the corresponding decay constants,
which have moderate deviations from phenomenological values. For the decay constant
of $\pi(1300)$ the experimental upper bound of 8.4 MeV \cite{Diehl:2001xe} is
easily respected and the results appear to be in an interesting ballpark
when compared to Ref.~\cite{Maltman:2001gc}. The results for the decay constant
of the lightest axial vector meson are likewise broadly consistent with
phenomenological results as obtained in \cite{Yang:2007zt} using light-cone sum rules.

Pion form factors have been studied in \cite{Kwee:2007dd} for HW and SW models with the conclusion that the former tend to work better.

\section{Holographic transition form factors and short distance constraints}\label{sec:TFFSDC}

In HLBL, the most important contributions come from exchanges of single
neutral mesons from the axial sector, which involve their
transition form factors, i.e., their coupling to two (real or virtual) photons.

To compute the transition form factor of an individual $C=+1$ meson one first looks at $\langle \phi(x) \rangle_{V_b}$ where $\phi_n$ is the corresponding 4d field that can be found in one of the towers being multiplied by its holographic wavefunction $\phi_n(z)$ and the subscript $V_b$ indicates that $V_{\mu}(x,z)$ now solves the equations of motion subject to the boundary condition $V(x,0)=V_b(x)$. In Fourier space this is solved using the bulk to boundary propagator by
\begin{equation}
    \tilde{V}_{\mu}(Q,z) = \sum_n \psi_n(z) \tilde{V}^{(n)}_{\mu}(Q)+ J(Q,z)\tilde{V}_{b,\mu}(Q),
\end{equation}
and the axial vector and pion fields only having the normalizable terms.

The TTFs of pseudoscalars and axial vectors come exclusively from the Chern-Simons action, which in form notation reads
\begin{align}\label{SCS}
    &S_{CS}= S_{CS}^L-S_{CS}^R, \nonumber \\
    &S_{CS}^B= \frac{N_c}{24 \pi^2} \int \text{tr}(BF^2-\frac{i}{2}B^3F-\frac{1}{10}B^5).
\end{align}

\begin{figure}[t]
\centering
\includegraphics[width=0.4\textwidth]{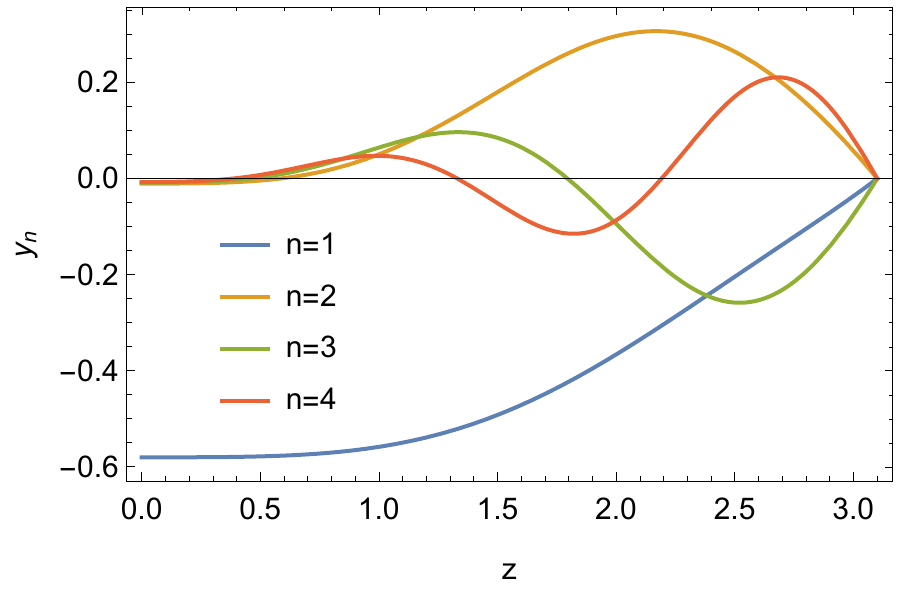}
\includegraphics[width=0.4\textwidth]{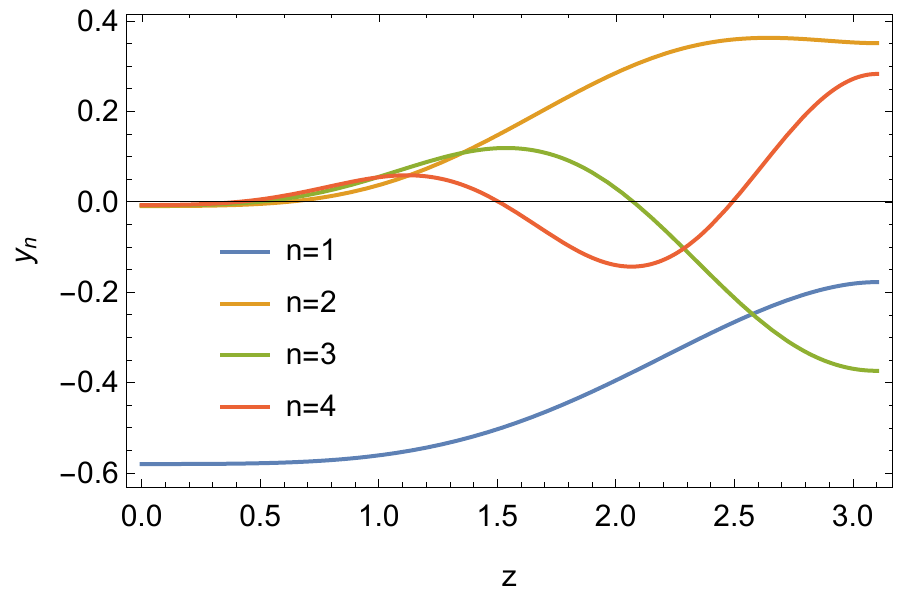}
\caption{The first four pion mode functions $y_n(z)$ in the HW1m model and in the HW3m version, with $z$ in units of inverse GeV; the $y_n$ have units of GeV, with $y_n(0)=-g_5 f_{\pi_n}$ The difference between these two models is the different IR boundary conditions. (Figure taken from \cite{Leutgeb:2021mpu})
\label{fig:ymodes}}
\end{figure}

In the HW1 case, where the axial vector field $A_M$ obeys the same IR boundary conditions as $V_M$, one needs to subtract an extra term localized at $z_0$ \cite{Grigoryan:2007wn,Leutgeb:2021mpu}. 
For the TFFs of the pions for example one finds
\begin{equation}
    \mathcal{F}_{\pi_n}(Q_1,Q_2)=- \frac{N_c g_5}{12\pi^2} \int_0^{z_0} dz \mathcal{J}(Q_1,z)\mathcal{J}(Q_2,z) \partial_z \phi_n 
\end{equation}
up to the possible IR subtraction term.
In figure \ref{fig:ymodes} the holographic wave functions of the ground-state and excited pions are plotted for different choices of IR boundary conditions in terms of $y_n(z)=\partial_z \phi_n/z$.
In the HW models with a scalar field $X$ it is possible to satisfy both constraints \eqref{eq:OPEconstr} and \eqref{eq:BrodskyLepconstr} for all pseudoscalar mesons and the analogous constraints for the TFFs of the axial vector mesons.

In fact, the bottom-up holographic models 
reproduce
exactly \cite{Grigoryan:2008up,Leutgeb:2019zpq} the asymptotic dependence on the asymmetry parameter
$w=(Q_1^2-Q_2^2)/(Q_1^2+Q_2^2)$ derived by Brodsky and Lepage in pQCD \cite{Lepage:1979zb,Lepage:1980fj,Brodsky:1981rp,Efremov:1979qk,Hoferichter:2020lap},
\bea
&&\mathcal{F}_{\pi_n}(Q_1,Q_2)\nonumber\\
&&\quad\to\frac{g_5^2 N_c}{12\pi^2}\frac{2 f_{\pi_n}}{Q_1^2+Q_2^2}
\left(\frac1{w^2}-\frac{1-w^2}{2w^3}\ln\frac{1+w}{1-w}\right)
\eea
at large $Q_i^2\to\infty$ (the correct prefactor is obtained in the HW1 and HW3 models,
where the SDCs can be implemented fully).

The analogous TFF in the top-down Sakai-Sugi\-moto model fails to satisfy these constraints, but that is to be expected since the gravitational theory is actually supposed to be dual to a five-dimensional field theory, which becomes four-dimensional large-$N_c$ QCD only at low energies below the Kaluza-Klein scale $M_\mathrm{KK}$.

In the chiral and large $N_c$ limit one can derive an exact expression for the divergence of the $VVA$ correlator. The Goldstone bosons of the broken $U(N_f)_A$ have an overlap with the axial current proportional to its decay constant. Polology then allows to derive \eqref{eq:lowenergyconstr}, thereby fixing the normalization of the TFF. Weak interactions couple to the axial current and $f_{\pi}$ appears in this way in weak processes and can be extracted from experiment.
Away from the chiral limit, but still in the isospin-symmetric limit the holographic models give the following interesting generalization \cite{Leutgeb:2021mpu} of \eqref{eq:lowenergyconstr}
\begin{equation}
    \sum_{n=1}^{\infty}f_{\pi_n}\mathcal{F}_n(0,0)=\frac{N_c}{12\pi^2}.
\end{equation}
For $n>1$ all $f_{\pi_n}$ are proportional to positive powers of the quark mass and vanish in the chiral limit, but the higher $\mathcal{F}_n(0,0)$ are still non-zero in the chiral limit. 

Axial vector mesons are more complicated, because their scattering amplitude with two virtual photons involves two asymmetric structure functions  \cite{Pascalutsa:2012pr,Roig:2019reh,Zanke:2021wiq}. In the holographic QCD models considered here, this
scattering amplitude is solely due to the Chern-Simons action (\ref{SCS}), which
gives
\cite{Leutgeb:2019gbz,Cappiello:2019hwh}
\bea\label{calMa}
\mathcal{M}_{\mathcal{A}^a\gamma^*\gamma*}&=&i\frac{N_c}{4\pi^2}\mathrm{tr}(\mathcal{Q}^2 t^a)\,\varepsilon_{(1)}^\mu \varepsilon_{(2)}^\nu
\varepsilon_{A}^{*\rho} \epsilon_{\mu\nu\rho\sigma}\\
&&\hspace*{-5mm}\times\left[q_{(2)}^\sigma Q_1^2 A_n(Q_1^2,Q_2^2)-q_{(1)}^\sigma Q_2^2 A_n(Q_2^2,Q_1^2)\right],\nonumber
\eea
where
\begin{align}\label{TFFAn} 
A_n(Q_1^2,Q_2^2) &= \frac{2g_5}{Q_1^2} \int_0^{z_0}\!\!\! dz \!\left[ \frac{d}{dz} \mathcal{J}(Q_1,z) \right]
\mathcal{J}(Q_2,z) \psi^A_n(z).
\end{align}
Because $\mathcal{J}(Q,z) \equiv 1$ when $Q^2=0$, 
the Landau-Yang theorem \cite{Landau:1948kw,Yang:1950rg}, which states that an axial vector meson cannot decay into two real photons, is automatically satisfied.

The asymptotic behavior of (\ref{TFFAn}) reads \cite{Leutgeb:2019gbz}
\bea
A_n(Q_1^2,Q_2^2) &\to& 
\frac{g_5^2 F^A_{n}}{Q^4}
\frac1{w^4}\biggl[
w(3-2w)\nonumber\\
&&\qquad +\frac12 (w+3)(1-w)\ln\frac{1-w}{1+w}
\biggr],
\eea
which agrees with the pQCD behavior that was derived only
recently in Ref.~\cite{Hoferichter:2020lap}.

\begin{figure}
\hspace*{-1.5cm}\includegraphics[width=0.55\textwidth]{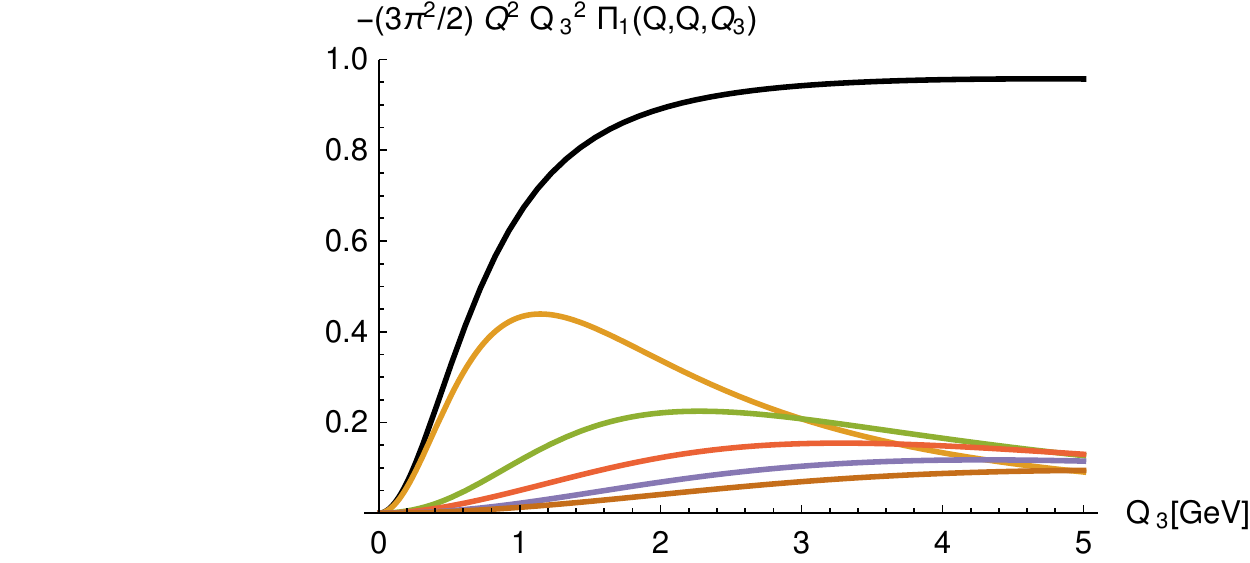}
\caption{Axial-vector contribution to
$Q_3^2 Q^2 \bar\Pi_1(Q,Q,Q_3)$ as a function of $Q_3$ at $Q=50$ GeV
in the HW2 model 
normalized to the asymptotic value \eqref{eq:MVLSDC} (with prefactor $g_5^2/(2\pi)^2$ set to one). 
The black line corresponds to the infinite sum over the tower of axial vector mesons, and the other lines
give the contributions of the 1st to 5th lightest axial vector mesons. (Figure taken from \cite{Leutgeb:2019gbz})
}
\label{fig:MV}
\end{figure}

As remarked at the end of section 3, single pseudoscalar or axial vector resonances cannot contribute to the leading short distance constraints on the HLBL tensor. In the holographic models infinite towers of resonances appear naturally and it turns out that it is the tower of axial vector mesons that is responsible for contributing to the two constraints \eqref{eq:MVConstr} and \eqref{eq:symlim}. The formula for the tower of axial vector meson contributions to $\bar{\Pi}_1$ in the asymmetric region is
\begin{align}
    &\bar{\Pi}_1(Q,Q,Q_3) \nonumber \\ &= -\frac{g_5^2}{2 \pi^4} \sum_{n=1}^{\infty} \int_0^{z_0} \;dz \left[\frac{d}{dz}\mathcal{J}(Q,z)\right]\mathcal{J}(Q,z)\psi^A_n(z) \nonumber \\ & \quad\times\frac{1}{(M_n^AQ_3)^2}\int_0^{z_0} \;dz' \left[\frac{d}{dz'}\mathcal{J}(Q_3,z')\right]\psi^A_n(z').
\end{align}
Resumming all the $n$-dependent terms and using the axial vector bulk-to-bulk propagator one can show for all HW models \cite{Leutgeb:2019gbz,Cappiello:2019hwh,Leutgeb:2021mpu} that in the same limit as in equation \eqref{eq:MVConstr} one gets 
\begin{equation}\label{eq:MVLSDC}
    \lim_{Q_3\rightarrow \infty}  \lim_{Q \rightarrow \infty} Q_3^2 Q^2 \bar{\Pi}_1(Q,Q,Q_3)= -\frac{g_5^2}{(2 \pi)^2}\frac{2}{3 \pi^2},
\end{equation}
which upon inserting the value for $g_5$ obtained from the asymptotics of the vector current correlator reproduces precisely the result obtained in QCD. 
In Fig.~\ref{fig:MV} the build-up of the correct asymptotic behavior is shown
for the HW2 model with $g_5=2\pi$ for $Q=50$ GeV.

As mentioned before, for large $N_c$ models in the chiral limit a stronger result holds, namely one can drop the left-most limit and  \eqref{eq:MVLSDC} remains valid for all $Q_3$. In the HW2 model without the fundamental scalar field $X$ one can indeed show analytically that this is indeed the case.

For the symmetric limit \eqref{eq:symlim} a similar computation also shows that axial vector mesons contribute to the limit, however this time they only reach $81$\% of the full value \cite{Cappiello:2019hwh,Leutgeb:2021mpu}. 
Thus in these holographic models it is the tower of axial vectors that is responsible for satisfying the  short distance constraints on the HLBL tensor at least qualitatively; the notorious MV-LSDC is satisfied exactly in the HW1 and HW3 models.

Summing 
the infinite tower of pseudoscalar mesons can also lead to a different asymptotic behavior
than given by the individual contributions.
In \cite{Leutgeb:2021mpu} it was shown that the massive HW models with $M_X^2=-3$ give
\begin{align}\label{Pi1piasymptotic}
-Q^2 Q_3^2 \bar\Pi_1^{(\pi)}(Q,Q,Q_3) 
\sim\frac{M_q^2}{6\pi^2}\frac{\ln(Q_3^2)}{Q_3^2}
\to 0,
\end{align}
which means that the tower of pions does not contribute to the leading short distance constraints. Also, \eqref{Pi1piasymptotic} is proportional to the quark mass indicating that the contributions of the massive pions are even more suppressed in the chiral limit. Note that in these models $M_q$ is a fixed parameter that does not run with energy.
In the HW models where $M_X^2$ is changed from its standard value, also a (fractional) power-law enhancement due to the resummation is possible, but never enough to change
the result that the LSDC is governed exclusively by the infinite tower of axial vector mesons. In fact, when $M_X^2<-3$, as is the case for the HW1m' and the HW3m' model that
fit the mass of the lightest axial vector meson and the first excited pseudoscalar, respectively (see Table \ref{tab:mdc}), even the logarithmic enhancement in (\ref{Pi1piasymptotic}) due to the resummation of massive pions disappears.

\section{Comparison of holographic transition form factors with experimental data}

In the HLBL contribution to $a_\mu$, the most important contribution is due
to the neutral pion, for which the singly-virtual TFF is well studied experimentally while direct data for the doubly-virtual case are still missing.

\begin{figure}
\includegraphics[width=0.48\textwidth]{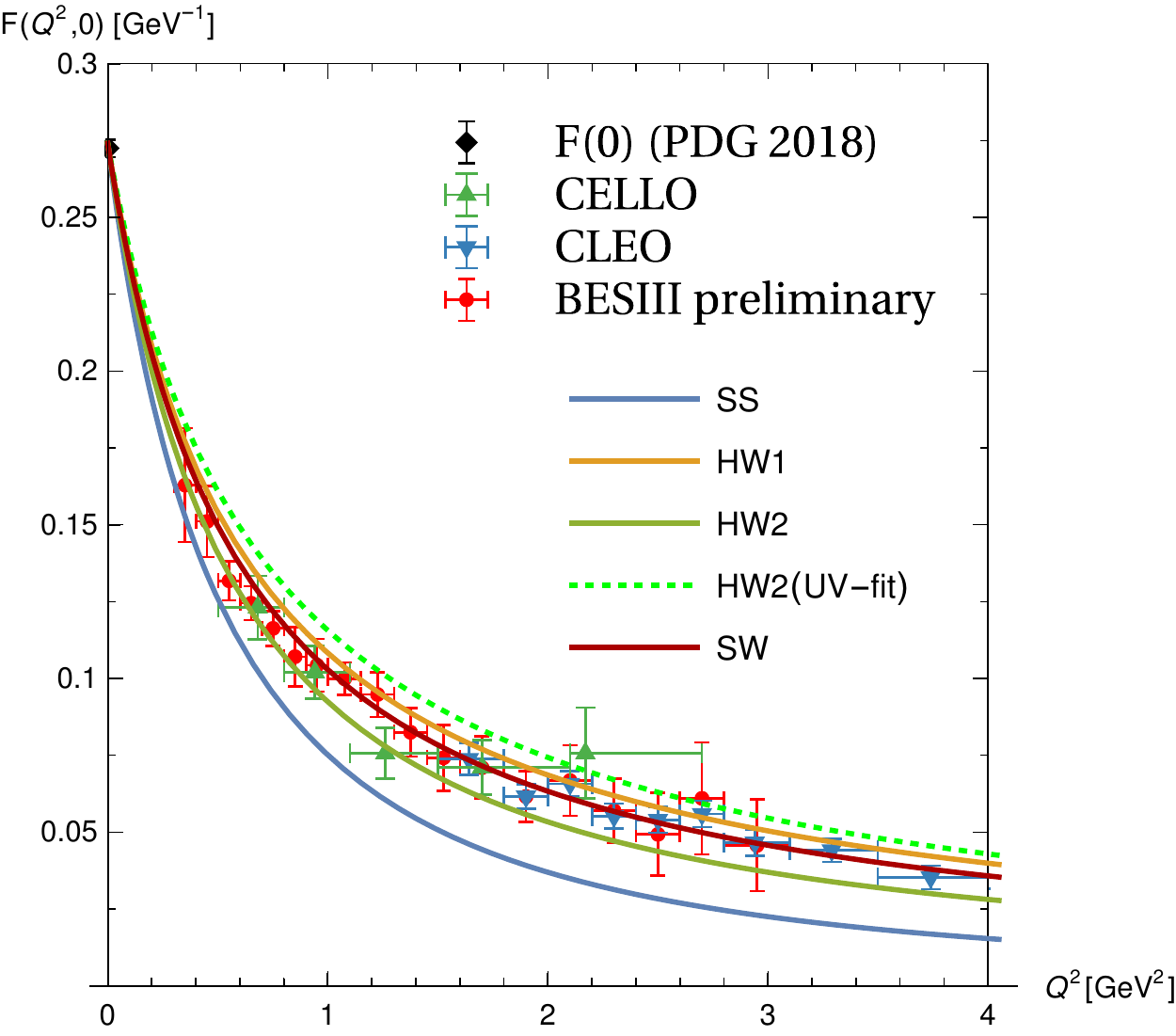}
\caption{Data for the $\pi^0$ TFF from CELLO, CLEO, BESIII-preliminary as compiled in Fig.~3 of Ref.~\cite{Danilkin:2019mhd} compared to the results of the various holographic models.
The result of the recent dispersion relation study of Ref.~\cite{Hoferichter:2018kwz}
(not shown to avoid overcrowding) 
lies right in between the SW (red) and HW1 (orange) result, with the lower end of the error band given in Ref.~\cite{Hoferichter:2018kwz}
nearly coinciding with the SW result. 
}
\label{fig3comp}
\end{figure}

In Fig.~\ref{fig3comp}
, the compilation of singly-virtual data of Ref.~\cite{Danilkin:2019mhd} are compared with the holographic results of
the chiral holographic models of Sakai and Sugimoto (SS), 
HW1, HW2, and a simplified soft-wall (SW) model. The HW1 and HW2 models, which attain
100\% and 62\% of the leading-order asymptotics, respectively, nicely bracket
the experimental results at all energies, while an optimal fit appears to take
place for the SW model which happens to produce 89\% of the SDC.
The SS model, which at high energies decays faster by an extra factor of $\sqrt{Q^2}$
\cite{Leutgeb:2019zpq}, is below the experimental data for $Q^2\gtrsim 0.5\, \mathrm{GeV}^2$.

In the doubly-virtual case, the bottom-up holographic models, in particular HW1 and SW, compare well \cite{Leutgeb:2019zpq}
with results from the data-driven dispersive approach of Ref.~\cite{Hoferichter:2018kwz}
as well as the recent lattice extrapolation of Ref.~\cite{Gerardin:2019vio}.
Because in the calculation of the HLBL contribution to $a_\mu$, the TFFs are needed
for both singly and doubly-virtual configurations, and previous models for the
pion TFF do not reproduce the asymptotic $w$-dependence, the holographic results
seem to offer an important improvement. (Recently, \cite{Danilkin:2019mhd} proposed
a new interpolating ansatz for the pseudoscalar TFFs which matches the asymptotic
pQCD result. However, for $Q^2\lesssim 2$ GeV the holographic pion TFF agrees significantly
better with the mentioned data-driven and the lattice results in the doubly virtual region \cite{Leutgeb:2019zpq}.)

For axial vector mesons, experimental information on the TFFs is rather limited.
The predictions of holographic QCD models, which seem to work surprisingly well
for pions after having fixed a minimal set of parameters, 
are therefore particularly interesting. The counterparts of the pseudoscalar
$\pi^0$, $\eta$, and $\eta'$
are the neutral $a_1(1260)$ and the isoscalars $f_1(1285)$ and $f_1'(1420)$.
Because the Landau-Yang theorem forbids decays into two real photons, 
one usually defines a so-called equivalent two-photon decay width through
\cite{Schuler:1997yw,Pascalutsa:2012pr}
\be
\tilde \Gamma_{\gamma\gamma}=\lim_{Q_1^2\to0} \Gamma(\mathcal A\to\gamma^*_L\gamma_T) M_A^2/(2Q_1^2),
\ee
which is determined by the value $A(0,0)$ in (\ref{calMa}). From the L3 experiment
there are data for $f_1(1285)$ and $f_1'(1420)$ \cite{Achard:2001uu,Achard:2007hm}. The former correspond to \cite{Zanke:2021wiq,Leutgeb:2021mpu}
\be
|A(0,0)|_{f_1(1285)}^\mathrm{exp.}=16.6(1.5)\,\text{GeV}^{-2}.
\ee
According to Refs.~\cite{Roig:2019reh,Masjuan:2020jsf} the corresponding value for the lightest $a_1$ meson
reads
\be
|A(0,0)|^\mathrm{exp}_{a_1(1260)}=19.3(5.0)\,\text{GeV}^{-2}. 
\ee
In the massive HW models one finds the
range \cite{Leutgeb:2021mpu}
\be|A(0,0)|_{n=1}=(19.95\ldots21.29)\,\text{GeV}^{-2},
\ee
which is compatible
with the latter, but above the value obtained for the $f_1(1285)$ meson; the HW2 model,
which has only 62\% of the leading-order pQCD asymptotics, yields a smaller value of 16.63 GeV$^{-2}$. Evidently, the holographic QCD models give at least a reasonable estimate of the ballpark.

\begin{figure}
\centering
\includegraphics[width=0.48\textwidth]{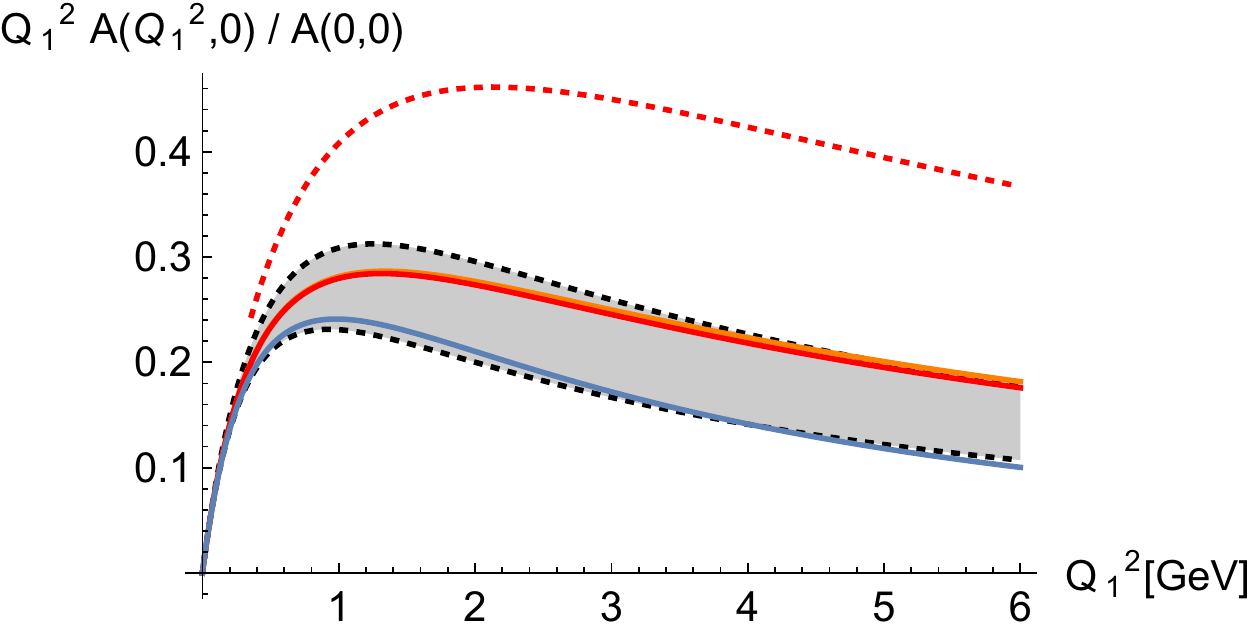}
\caption{Single-virtual axial vector TFF from holographic models 
(SS: blue, HW1: orange, HW2: red) compared with
dipole fit of L3 data for $f_1(1285)$ (grey band). The parameters
of all models are fixed by matching $f_\pi$ and $m_\rho$.
The results for HW1 and HW2 almost coincide, with HW2 at most a line
thickness above HW1. When the mass scale $z_0^{-1}$
is not fixed by $m_\rho$ but instead matched to the pQCD with $N_c=3$, HW2(UV-fit) instead gives
the significantly larger result denoted by the red dotted line. (Figure taken from \cite{Leutgeb:2019gbz})}
\label{fig:f1ffsingle}
\end{figure}
\begin{figure}
\centering
\includegraphics[width=0.45\textwidth]{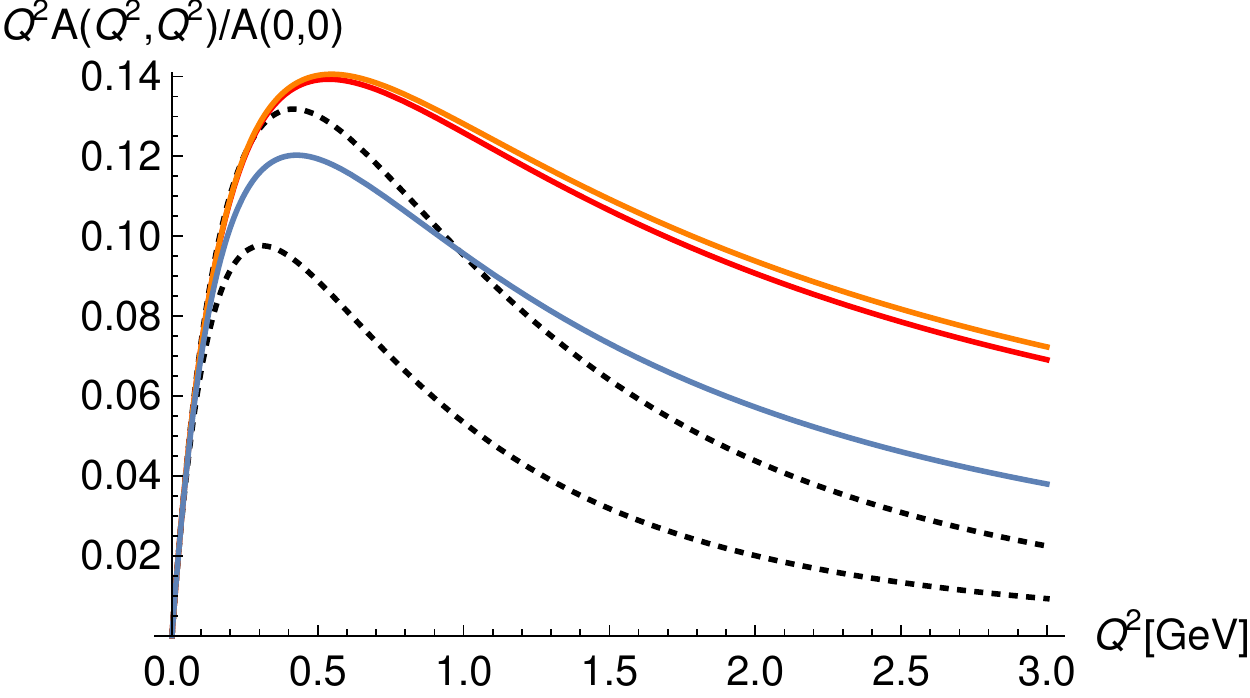}
\caption{Double-virtual axial vector TFF for $Q_1^2=Q_2^2=Q^2$ from holographic models 
(SS: blue, HW1: orange, HW2: red). The black dashed lines denote the extrapolation
of L3 data with a dipole model for each virtuality as used in the calculation of $a_\mu^{f_1}$ in Ref.~\cite{Pauk:2014rta}. (Figure taken from \cite{Leutgeb:2019gbz})}
\label{fig:f1ffsym}
\end{figure}

The experimental results for the singly-virtual TFF for $f_1(1285)$ \cite{Achard:2001uu}
have been found to be close to a dipole form, which is shown in Fig.~\ref{fig:f1ffsingle}
as a grey band together with the results for the HW1 and HW2 models and the SS model.
The former two almost coincide when divided by $A(0,0)$, and the result of the SS model
is significantly smaller, but all of them are compatible with the experimental result.
In the important region $Q_1^2\lesssim 2$ GeV$^2$, the HW results are however in better agreement.

In previous evaluations of the axial vector contribution to $a_\mu$, mostly a simple dipole ansatz has been chosen for all virtualities, in particular in Ref.~\cite{Pauk:2014rta} used in the WP \cite{Aoyama:2020ynm}. Fig.~\ref{fig:f1ffsym} shows the difference of the doubly-virtual TFF at $Q_1^2=Q_2^2$ obtained in the holographic models compared to such a dipole ansatz with the parameters obtained from the L3 experiment, which suggests that a dipole ansatz likely underestimates the contributions of the individual contributions of axial vector mesons. However, as the discussion in Sect.~\ref{sec:TFFSDC} has shown, the whole tower of axial vector mesons needs to be considered to assess the numerical importance of the LSDC in the HLBL contribution to $a_{\mu}$.
 
\begin{figure}
\centering
\includegraphics[width=0.48\textwidth]{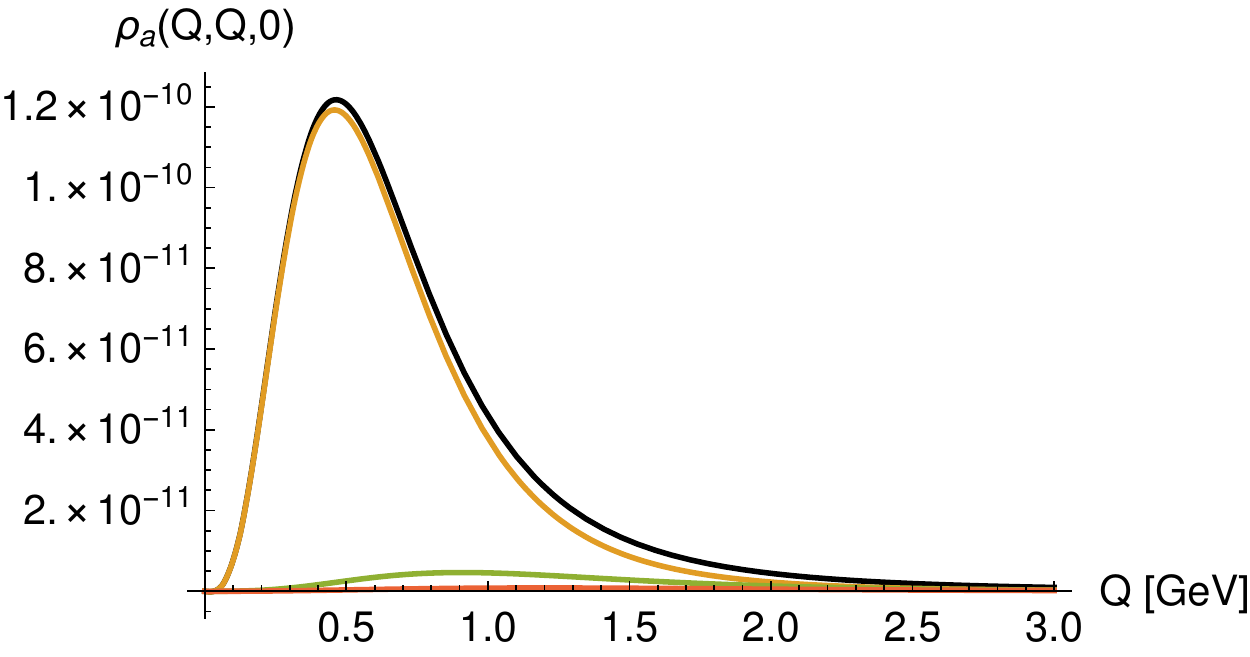}
\caption{The integrand $\rho_a(Q_1,Q_2,\tau)$ in (\ref{eq:integral}) in units of GeV$^{-2}$
for $Q_1=Q_2$ and $\tau=0$ (implying $Q_3=\sqrt{2}Q$) in case of the HW2 model. The black line is
the result from the infinite sum over the tower of axial vector mesons, the other lines
give the contributions of the 1st to 3rd lightest axial vector meson multiplets.
(Figure taken from \cite{Leutgeb:2019gbz})}
\label{fig:rhoaAV}
\end{figure}

\section{Results for the HLBL contribution to $a_{\mu}$}
 
 
In order to obtain the HLBL contribution to the muon anomalous magnetic moment, the various form factors have to be used for the respective components of the HLBL scattering amplitude analogous to \eqref{eq:picontr} and inserted into the master formula \eqref{eq:integral}. 

For the ground-state pseudoscalar bosons, such a calculation has been carried out first in \cite{Hong:2009zw} for the massive HW1 model, but using only the first few terms of a mode expansion of the bulk-to-boundary operator, and for a set of chiral holographic QCD models (SS, HW1, HW2, and SW) in \cite{Cappiello:2010uy} by replacing their complicated form factors
with simplified versions. A complete evaluation of the models considered in \cite{Cappiello:2010uy} was only recently carried out in \cite{Leutgeb:2019zpq}, yielding the results given in table \ref{tab:amuGB}.

\begin{table}
\caption{Results in multiples of $10^{-11}$ for $f_\pi=92.4 \,\text{MeV}$ for the Sakai-Sugimoto model and the various chiral bottom-up models \cite{Leutgeb:2019zpq}. 
For estimating also the contributions $a_\mu^{\eta}$ and $a_\mu^{\eta'}$ 
$F(0,0)$ was rescaled by the central experimental values quoted in \cite{Danilkin:2019mhd}.
For $\eta'$ a second value is given which includes a
presumably more realistic extrapolation obtained by additionally upscaling the mass scale within $F(Q_1^2,Q_2^2)$
by +10\% in line with the higher $\Lambda$ parameter in the fits carried out in \cite{Danilkin:2019mhd}.
HW2 and HW2$_\mathrm{UV-fit}$ with the lower values for $\eta'$ correspond almost exactly
to the parameter choices ``Set 1'' and ``Set 2'' of Cappiello et al.\ \cite{Cappiello:2019hwh}.
}
\label{tab:amuGB}       
\centering
\begin{tabular}{lcccr}
\hline\noalign{\smallskip} 
   & $a_\mu^{\pi^0}$ & $a_\mu^{\eta}$ & $a_\mu^{\eta'}$ & sum \\ 
\noalign{\smallskip}\hline\noalign{\smallskip}
SS  & 48.3 & 11.7 & 7.8$|$9.5 & 67.8$|$69.4 \\
SW  & 59.2 & 15.9 & 11.2$|$13.4 & 86.3$|$88.5 \\
HW1 & {65.2} & 18.2 & 13.2$|$15.6 & 96.6$|${99.0} \\
HW2  & 56.6 & 14.8 & 10.3$|$12.4 & 81.7$|$83.7 \\
HW2$_\text{UV-fit}$\hspace*{-5mm} & 75.4 & 21.9 & 16.1$|$19.0 &  113.4$|$116.3 \\
WP \cite{Aoyama:2020ynm} & 63.6(2.7) & 16.3(1.4) & 14.5(1.9) & 94.3(5.3) \\
\noalign{\smallskip}\hline 
\end{tabular}
\end{table}

\begin{table}
\caption{The contribution of the infinite tower of axial vector mesons to $a_\mu^\mathrm{AV}$ in multiples of $10^{-11}$ obtained in Ref.~\cite{Leutgeb:2019gbz}. The entries $j\le n$ give the contribution of the first $n$
axial vector multiplets.
}
\centering
\begin{tabular}{lcccccc}
\hline\noalign{\smallskip}
 & $j=1$ & $j\le2$ & $j\le3$ & $j\le4$ & $j\le5$ & $a_\mu^\mathrm{AV}$ \\
\noalign{\smallskip}\hline\noalign{\smallskip}
HW1 & 31.4 & 36.2 & 37.9 & 39.1 & 39.6 & 40.6  \\
HW2 & 23.0 & 26.2 & 27.4 & 27.9 & 28.2 & 28.7  \\
HW2$_\text{UV-fit}$\hspace*{-3mm}& 23.7 & 26.9 & 28.1 & 28.6 &  28.9 & 29.4  \\
\noalign{\smallskip}\hline
\end{tabular}
\label{tab:amuj}
\end{table}

\begin{table}
\caption{Combined results for the HW1 and HW2 models, with the axial vector contribution split into longitudinal and transverse contributions}
\centering
\begin{tabular}{lcc}
\hline\noalign{\smallskip}
 & HW1 & HW2 \\
\noalign{\smallskip}\hline\noalign{\smallskip}
$a_\mu^{[\pi^0+\eta+\eta']}$ & 99.0 
& 83.7 
\\
$a_\mu^\mathrm{AV}[L+T]$ & 40.6 [23.2+17.4] & 28.7 [16.6+12.0]  \\
\noalign{\smallskip}\hline\noalign{\smallskip}
$a_\mu^{\pi^0+\eta+\eta'+\mathrm{AV}}$ & {140} & 112  \\
\noalign{\smallskip}\hline
\end{tabular}
\label{table4}
\end{table}

While for pseudoscalar bosons, only $\bar\Pi_{1\ldots3}$ contribute,
axial vector meson contributions involve all $\bar\Pi_i$ functions in \eqref{eq:integral}. This was evaluated
in Ref.~\cite{Leutgeb:2019gbz} using the chiral HW1 and HW2 models with the results given in Table \ref{tab:amuj}. As shown in Fig.~\ref{fig:rhoaAV},
the integrand of the axial vector contributions is strongly dominated by the lightest axial vector meson, but the higher excitations cannot be neglected. As seen from Table \ref{tab:amuj}, they increase the axial vector contribution by 29\% in the HW1 model and by 25\% in the HW2 model.
In Ref.~\cite{Cappiello:2019hwh} the analogous calculation was carried out for two sets of parameters in the HW2 model,
where ``Set 1'' corresponds essentially to the choice made in \cite{Leutgeb:2019gbz}, namely to fit the infrared parameters $f_\pi$ and $m_\rho$, which leads to an incomplete fit (62\%) of the UV asymptotics;
``Set 2'' instead corresponds to what has been called HW2(UV-fit) above, where one has full UV asymptotics but an excessively heavy rho meson with $m_\rho=987$ MeV. The latter has the effect that the pseudoscalar contribution is overestimated, while the axial vector meson contribution remains almost unchanged.\footnote{The fact that the higher axial vector meson modes do not become more important for HW2(UV-fit) has to do with the relatively large masses of excited
axial vector mesons in this case ($m_{a_1^*}^\text{HW1}=2154$ MeV  and $m_{a_1^*}^\text{HW2}=2261$ MeV but
$m_{a_1^*}^\text{HW2(UV-fit)}=2880$ MeV \cite{Leutgeb:2019gbz}). Note that the mass of the first excited $a_1$ meson according to PDG \cite{PDG20}
is only 1655(16) MeV.}

\begin{figure}
\centering
\includegraphics[width=0.45\textwidth]{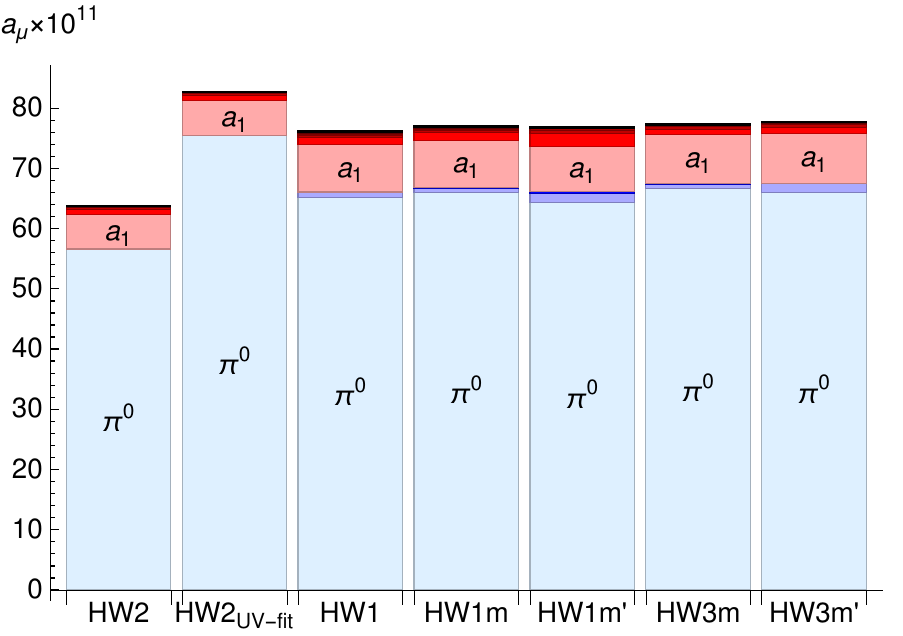}
\caption{Bar chart of the individual contributions to $a_\mu^{\pi\cup a_1}$
in the various HW models, with excited modes given by increasingly darker colors, blue for the $\pi^0$'s, red for the $a_1$'s.}
\label{fig:barchart}
\end{figure}

In \cite{Leutgeb:2019zpq,Leutgeb:2019gbz,Cappiello:2019hwh}, the chiral HW models were used for the TFF but in the propagator of the pion the physical mass was inserted by hand.
In Ref.~\cite{Leutgeb:2021mpu} the HW1/HW3 models were studied with finite quark masses in the isovector channels, and also the effects of the infinite tower of excited pions was evaluated. The results are shown
graphically in Fig.~\ref{fig:barchart}. The massive HW models have the
advantage that $f_\pi$ and $m_\rho$ can be kept at physical values while
100\% of the UV asymptotics in the TFFs and in the MV-LSDC are attained.

However, as argued in \cite{Leutgeb:2021mpu}, it may be more realistic
to demand only $\sim 90\%$ of the correct leading-order UV asymptotics, because at moderately large energies next-to-leading order corrections of the UV behavior of the corresponding magnitude are present \cite{Melic:2002ij,Bijnens:2021jqo}. The maximum of pristine HW1/HW3 results
could therefore be viewed as upper limits, and estimates for lower
limits were proposed by tuning $g_5^2$ such that $85\%$ of the SDCs are obtained. This leads to
\be
a_\mu^{\pi^0}=(60.5\ldots66.6)\times 10^{-11},
\ee
which nicely brackets the WP result \cite{Aoyama:2020ynm} of $63.6(2.7)\times 10^{-11}$.
A flavor symmetric extension of the results for the contributions
from the infinite towers of axial vector mesons and pseudoscalars leads to
the estimate \cite{Leutgeb:2021mpu}
\bea\label{extendedranges}
&&a_\mu^{\mathrm{P}^*}\equiv 4a_\mu^{\pi^*}=(3.2\ldots7.2)\times 10^{-11}, \nonumber\\
&&a_{\mu(L)}^{\mathrm{A}}=(20.8\ldots 25.0)\times 10^{-11} ,\nonumber\\
&&a_\mu^{\mathrm{A}}=(36.6\ldots 43.3)\times 10^{-11} ,\nonumber\\
&&a_\mu^{\mathrm{A}+\mathrm{P}^*}=(39.8\ldots 50.5)\times 10^{-11}.
\eea

The latter results (with or without\footnote{While the decay constants of the excited pseudoscalars in the HW1/HW3 models are found to have values compatible with experiment, as discussed above, the coupling of the first excited pseudoscalar to photons appears to be significantly above the upper bound obtained in \cite{Colangelo:2019uex} for $\pi(1300)$.} the excited pseudoscalar contributions) could be compared to the WP \cite{Aoyama:2020ynm} values attributed to the axial sector and contributions related to the SDC,
$a_\mu^\mathrm{WP,axials}=6(6)\times 10^{-11}$ and $a_\mu^\mathrm{WP,SDC} =15(10)\times 10^{-11}$, which with linearly
added errors gives $21(16)\times 10^{-11}$, which is significantly smaller.
While the estimates for the SDC contribution is more or less compatible
\cite{Ludtke:2020moa,Colangelo:2021nkr}, the main difference comes from the (transverse) contribution of axial vector mesons (Table \ref{table4}). The holographic QCD
results thus strongly suggest that their contributions have been underestimated so far. In fact, a recent re-evaluation of predictions
from resonance chiral models also suggests a larger contribution
from axial vector mesons \cite{Masjuan:2020jsf}.

Recently, also the contribution of scalar mesons has been worked out \cite{Cappiello:2021vzi}
in a variant of the HW model where the scalar $X$ has additional
interaction terms $|X|^2 F^2$ which can be fitted to match phenomenological
input on the decay widths of $f_0(500)$, $f_0(990)$, and the $a_0(980)$, albeit by using a different set of parameters for each of them. Here the holographic result for the sum of their contribution to $a_\mu$ is
\be
a_\mu^\mathrm{scalar}=-9(2)\times 10^{-11},
\ee
in perfect agreement with a recent evaluation within the dispersive approach \cite{Danilkin:2021icn}.

\section{Conclusion and Outlook}
 
The simplest (HW) models of QCD where the bulk geometry is AdS and thus does not account for a running coupling constant and where conformal symmetry is broken by a sharp cut-off have nevertheless proven to
offer valuable insight in how the long-standing difficulty of hadronic
models to account for the MV-LSDC can be resolved while being compatible with the chiral limit. Moreover, with a minimal set of parameters to fix,
quantitative predictions frequently turn out to reproduce experimental
and phenomenological results with errors to be expected from a large-$N$ approximation to QCD, although no top-down construction for a full holographic dual to (large-$N$) QCD is available. In cases where high precision is required, as in the case of the HVP contribution to the
muon anomalous magnetic moment, the simple holographic models considered cannot help to resolve the uncertainties of the SM prediction resulting
from the recent discrepancy of data-driven and lattice results. However, in the case of the HLBL contribution, where the theoretical error is currently at the level of 20\%, and where in particular the important contribution of axial vector mesons has an uncertainty of 100\%, holographic QCD can certainly provide valuable clues, complementary
to ongoing efforts with data-driven approaches (e.g., \cite{Zanke:2021wiq}) and all-inclusive lattice evaluations \cite{Blum:2019ugy,Chao:2021tvp}.

It is certainly of interest to extend further the existing holographic calculations of various HLBL contributions to $a_\mu$. On the one hand, additional hadronic channels can be explored,
in particular scalar and tensor mesons as well as glueballs, where experimental data are sparse or lacking. 
On the other hand, the holographic QCD models reviewed here can be replaced by refined models which take into account the mass of strange quarks and the Witten-Veneziano mass from the U(1)$_A$ anomaly. Moreover, the simple AdS background could be replaced by one that represents better the behavior of hadronic observables at higher energies, which is indeed achieved in the various models of improved holographic mentioned above.

\begin{acknowledgements}
J.L.\ and J.M.\ have been supported by the FWF doctoral program
Particles \& Interactions, project no.\ W1252-N27;
J.M.\ has been supported also by FWF project no.\ P33655. 
\end{acknowledgements}

\bibliographystyle{JHEP}
\bibliography{hlbl}   


\end{document}